\documentclass[graybox]{svmult}

\usepackage{type1cm}        
\usepackage{makeidx}         
\usepackage{graphicx}        
\usepackage{multicol}        
\usepackage[bottom]{footmisc}

\usepackage{newtxtext}       %
\usepackage[varvw]{newtxmath}       

\usepackage{bm}

\usepackage[square,comma,numbers,sort&compress]{natbib}
\usepackage[colorlinks=true,linkcolor=blue,citecolor=red]{hyperref}%
\usepackage{color}

\def\atan{\mathrm{atan}}

\def\x{\bm{x}}

\def\R{\mathbb{R}}
\def\pa{\partial\Omega}

\def\I{\mathbb{I}}

\def\E{\mathbb{E}}

\makeindex             

\begin{document}

\title*{Boundary homogenization for target search problems}

\author{Denis S. Grebenkov and Alexei T. Skvortsov}
\institute{
Denis S. Grebenkov \at 
Laboratoire de Physique de la Mati\`{e}re Condens\'{e}e,
CNRS -- Ecole Polytechnique, Institut Polytechnique de Paris, 91120 Palaiseau, France, 
\email{denis.grebenkov@polytechnique.edu}\and
Alexei T. Skvortsov \at 
Maritime Division, Defence Science and Technology Group, Melbourne,
VIC 3207, Australia,
\email{alexei.skvortsov@defence.gov.au}}

%
\maketitle

\abstract{
In this review, we describe several approximations in the theory of
Laplacian transport near complex or heterogeneously reactive
boundaries.  This phenomenon, governed by the Laplace operator, is
ubiquitous in fields as diverse as chemical physics, hydrodynamics,
electrochemistry, heat transfer, wave propagation, self-organization,
biophysics, and target search.  We overview the mathematical basis and
various applications of the effective medium approximation and the
related boundary homogenization when a complex heterogeneous boundary
is replaced by an effective much simpler boundary.  We also discuss
the constant-flux approximation, the Fick-Jacobs equation, and other
mathematical tools for studying the statistics of first-passage times
to a target.  Numerous examples and illustrations are provided to
highlight the advantages and limitations of these approaches.}

\section{Introduction}
\label{sec:intro}

The diffusion of Brownian particles near complex boundaries is
ubiquitous for many natural phenomena and engineering applications.
Common examples are heterogeneous catalysis, diffusion-limited
aggregation, tracer dispersion over complex canopies in atmosphere and
ocean, cell communication via chemical signals, gas exchange in the
human lungs and placentas, and many others
\cite{ben-Avraham_2010,Schuss_2013,Bejan_2000,Rice_1985,Hughes_1995,Krapivsky_2010,Lindenberg_2019,Lauffenburger_1993,Reva_2021,Filoche_2008,Nair_2007,Nepf_2007,Edburg_2010,Ramon_2013,Felici_2004,Grebenkov_2005,Serov_2016,Witten_1981}.
The mathematics of this process remains unchanged even if the
diffusive quantity is continuous (e.g., temperature profile, velocity
field, wave amplitude, magnetic field, vorticity) and thus not
directly related to any physical particle or Lagrangian marker.  In
this regard, diffusion is connected to the broad family of phenomena,
which are collectively referred to as \textit{Laplacian transport}.
The unified analytical framework of the underlying Laplace equation,
which also includes the diffusion equation at the steady state, is
thus applicable in fields as diverse as chemical physics, acoustics,
electrostatics, fluid dynamics, electrochemistry, wetting,
etc. \cite{Sapoval_1994,Filoche_2000,Levitz_2006,Grebenkov_2006,Bazant_2016,Rothstein_2010,Hewett_2016,Martin_2022,Fyrillas_2001,Blyth_2003,Crowdy_2011,Martin_2022a}.

The analytical results for diffusive transport near inhomogeneous
boundaries are readily available only for simple shapes (plane,
sphere, cylinder, etc.).  More complex boundary profiles require the
application of advanced analytical methods (e.g., conformal mapping,
eigenfunction decomposition, perturbation theory, scaling arguments,
etc.) and often lead to extensive numerical simulations via Monte
Carlo, finite element, or Lagrangian methods.  The aim of this review
is to show how the conventional framework of the effective medium
approximation and boundary homogenization can be applied to target
search problems to enable analytical progress even for boundaries of
rather complex profiles.  We show that this approach gives access not
only to some ``averaged'' properties of the system (e.g., the total
flux) but also to the first-passage time (FPT) statistics.  The latter
plays the central role in target search problems
\cite{Redner_2001,Metzler_2014,Stone_2016,Masoliver_2018} as the proxy
of efficiency of search strategies
\cite{Benichou_2011,Benichou_2014,Bressloff_2013}.

Generally speaking, the complexity of the boundary in the context of
diffusive transport can be attributed to two independent factors: (i)
the geometrical complexity of the boundary profile (spiky, fractal,
patchy, etc), and (ii) the distribution of the absorbing properties on
that boundary.  For instance, some parts of the boundary can be
absorbing and the other be reflecting.  This pattern can be rather
complex even if the geometrical profile of the boundary is simple
(e.g., flat boundary).  The present review deals with both kinds of
boundary complexity.  Instead of solving the target search problem for
a given complex boundary, boundary homogenization aims at replacing
such a boundary by an {\it equivalent simpler one}, for which the
target search problem admits an exact simple solution.  The review is
thus focused on the fundamental question of constructing such an
equivalent boundary, and provides theoretical answers and practical
ready-to-use recipes.

We begin with an illustrative example.  Consider particles diffusing
in a layer $0 < y < H$ between two parallel planes separated by
distance $H$.  At steady state, the particle concentration $C$ within
the layer obeys the Laplace equation
\begin{equation}
\label{ER01}
\Delta C = 0,
\end{equation}
where $\Delta = \partial^2/\partial x^2 + \partial^2/\partial y^2
+\partial^2/\partial z^2$ is the Laplace operator.  Assume that the
bottom plane $y =0$ is absorbing (i.e., any particle hitting this
plane disappears so that $C =0$ at $y = 0$), and the concentration of
the particles at the top plane $y=H$ is kept constant: $C=C_0$ (i.e.,
it is a permanent source of particles).  One can also think of $C$ as
a temperature profile inside a layer, whose two boundaries are kept at
constant temperatures $0$ and $C_0$.  The solution of this trivial
problem yields the linear profile of particle concentration
\begin{equation}
\label{ER1}
C  = \frac{C_0}{H} y = \frac{j_0}{D} y,
\end{equation}
where $j_0 = D C_0/H$ is the particle flux density, and $D$ is the
particle diffusivity.

\begin{figure}[t!]
\centering
\includegraphics[width=0.32\textwidth]{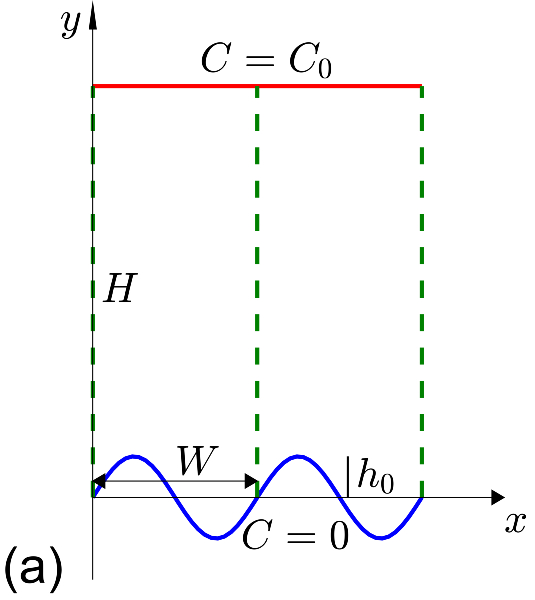} 
\includegraphics[width=0.32\textwidth]{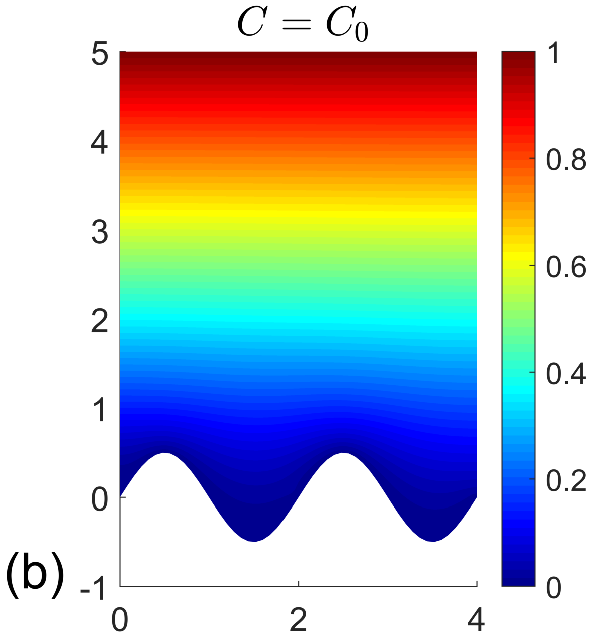} 
\includegraphics[width=0.32\textwidth]{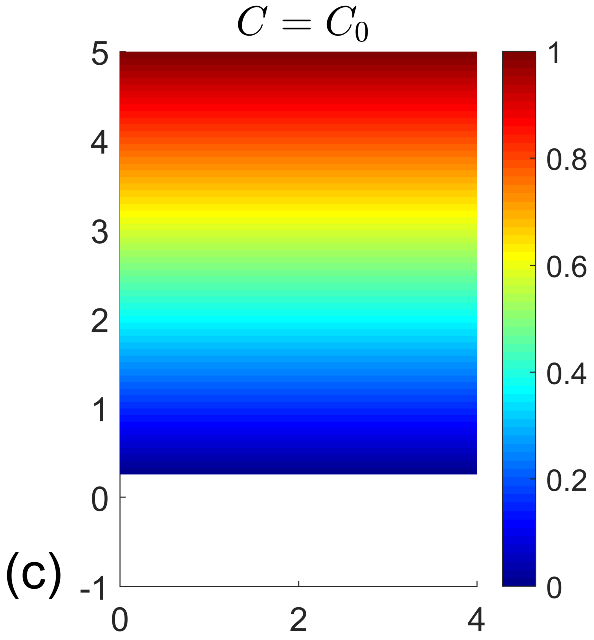} 
\caption{ 
{\bf (a)} A schematic view of Laplacian transport from the top (flat)
source at $y = H$ (in red) towards an absorbing boundary with
sinusoidal profile (in blue), with the period (or wavelength) $W$ and
amplitude $h_0$.  Dashed vertical lines delimit the periodic cell that
is repeated along $x$ coordinate.  The system is translationally
invariant along $z$ coordinate, which is perpendicular to the figure.
{\bf (b)} A numerical solution of the Laplace equation inside this
domain, with reflecting boundary conditions instead of periodic ones
on two vertical segments.  This numerical solution (obtained by a
finite-element method via PDEtool in Matlab) determines the diffusive
flux $J$ onto the absorbing boundary, which in turn sets via
Eq. (\ref{ER3}) the offset $\lambda$ (e.g., we get $\lambda \approx
0.26$ for $H = 5$, $W = 2$, and $h_0 = 0.5$).  {\bf (c)} The solution
$C = C_0 (y-\lambda)/(H-\lambda)$ in a simpler case with a flat
absorbing boundary at $y_0 = \lambda$, which yields the same diffusive
flux. }
\label{fig:sine}
\end{figure}

Next we assume that the bottom boundary is subject to a periodic
deformation with the period $W \ll H$ (Fig. \ref{fig:sine}a).  Near
the deformed boundary, the particle concentration is strongly
non-uniform, but for $y \gg W$ the transverse components of the
particle flux rapidly disappear and the concentration profile takes
the linear form (Fig. \ref{fig:sine}b):
\begin{equation} 
\label{ER2}
C  \approx \frac{C_0}{H - \lambda} (y - \lambda),
\end{equation}
where the offset parameter $\lambda$ has the dimension of length and
can be positive or negative (note that another convention with
$-\lambda$ instead of $\lambda$ in Eq. (\ref{ER2}) is often used).
This parameter aggregates all information about the boundary profile
and inhomogeneity of its absorbing properties that is relevant at far
field $y \gg W$.  For instance, in terms of $\lambda$, the diffusive
flux through the system is given by a simple expression
\begin{equation} 
\label{ER3}
 \frac{J}{J_0} = \frac{1}{1 - \lambda /H},
\end{equation}
where $J_0 = j_0 {\mathcal A}_0 = D C_0 {\mathcal A}_0/H$ is the
diffusive flux to a flat absorbing boundary, and ${\mathcal A}_0$ is
the cross-sectional area of the periodic cell.  At the same time, this
relation allows one to determine the offset $\lambda$ from the
diffusive flux $J$, which can be found numerically by solving the
Laplace equation or via Monte Carlo simulations:
\begin{equation}  \label{eq:lambda_def}
\lambda = H (1 - J_0/J).
\end{equation}

Similarly, the far-field solution (\ref{ER2}) remains valid for a flat
bottom boundary with absorbing and reflecting parts (see examples
below).  If the bottom boundary is fully reflecting, the concentration
$C$ is constant, and the diffusive flux $J$ is zero.  This situation
corresponds to the limit $\lambda = - \infty$.  Adding more and more
absorbing spots onto the bottom boundary increases the diffusive flux
and thus the offset $\lambda$, which thus ranges from $-\infty$ (flat
reflecting boundary) to $0$ (flat absorbing boundary).  Moreover, if
the bottom boundary is both deformed and heterogeneous, the far-field
solution may correspond to positive $\lambda$.

The linear form (\ref{ER2}) provides a simple interpretation for the
offset $\lambda$: the condition $y = \lambda$ determines the position
of the equivalent absorbing boundary that provides the same diffusive
flux through the system.  The goal of boundary homogenization is to
relate $\lambda$ to the geometrical properties of the original complex
boundary.  As this problem is relevant for various disciplines, the
aggregated parameter $\lambda$ bears different names: slip length in
the context of superhydrophobic coating \cite{Crowdy_2011}, blockage
coefficient in fluid dynamics \cite{Crowdy_2011,Martin_2022}, grid
parameter in electrostatics \cite{Hewett_2016}, displacement length in
heat transfer \cite{Fyrillas_2001,Blyth_2003}.  The mathematical
equivalence between steady-state diffusion and electrostatics, both
described by the Laplace equation, is particularly helpful for
bringing intuitive interpretations \cite{Sapoval_1994,Sapoval_1996}.
In fact, if $C_0$ is understood as the applied electric potential
(voltage) between two metal electrodes located at the top and at the
bottom, then $J$ becomes the electric current, while $Z = C_0/J$ is
the electric resistance (or impedance) of this system.  For instance,
if the bottom electrode is flat, one retrieves the classical formula
$Z_0 = C_0/J_0 = H/(D\mathcal{A}_0)$ for the resistance of a metal
wire of length $H$ and cross-sectional area $\mathcal{A}_0$, while
$1/D$ plays the role of electric resistivity.  As a consequence,
Eq. (\ref{ER3}) provides another interpretation of the offset
$\lambda$ as the relative difference between the electric resistances
$Z$ and $Z_0$:
\begin{equation} 
\label{ER3b}
 \frac{Z_0 - Z}{Z_0} = \frac{\lambda}{H} .
\end{equation}
Moreover, common rules for computing the overall resistance of an
electric circuit, composed of sequential and/or parallel connections
between its elements, can shed light onto various expressions for the
offset $\lambda$ that will be discussed below.

In chemical physics, it is more conventional to relate the offset
$\lambda$ to the trapping parameter $\kappa$ that appears in the Robin
(or radiation) boundary condition on a partially absorbing (reactive)
flat boundary located at some height $y_0$:
\begin{equation}
\label{ER4}
- D \frac{\partial C}{\partial n}  = \kappa C,
\end{equation}
where $\partial /{\partial n}$ is the normal derivative oriented
outwards the domain \cite{Collins_1949}.  The trapping parameter
$\kappa$, which has units of meter per second, is also known as
surface reactivity, forward reaction constant, permeability,
relaxivity, or the inverse of surface resistivity, depending on the
application field.  It can be checked by direct substitution of
Eq. (\ref{ER2}) into Eq. (\ref{ER4}) that these approaches are
equivalent and
\begin{equation}
\label{IE004}
\kappa = \frac{D}{y_0 - \lambda}.
\end{equation}
While the offset $\lambda$ is uniquely determined by the
shape of the bottom boundary and the inhomogeneity of its absorbing
parts, the value of the trapping parameter $\kappa$ depends on the
{\it choice} of the location $y_0$ of the effective flat boundary.
This ambiguity actually offers some flexibility for boundary
homogenization.  From the mathematical point of view, the trapping
parameter $\kappa$ should be restricted to nonnegative values, with
two relevant limits, $\kappa \rightarrow 0$ and $\kappa \rightarrow
\infty$, corresponding to the perfectly reflecting and perfectly
absorbing boundary, respectively.  When $\lambda < 0$, one can
naturally choose $y_0 = 0$.  However, if $\lambda > 0$, it is
mandatory to choose $y_0 \geq \lambda$ (note that the choice $y_0 =
\lambda$ corresponds to an absorbing boundary with $\kappa = \infty$,
as discussed above).

For simplicity of notations and more compact formulas, we will employ
both parameters $\lambda$ and $\kappa$ interchangeably, keeping in
mind that they are related via Eq. (\ref{IE004}).  One of the aims of
this review is to transfer the analytical results from the different
areas of Laplacian transport to the context of particle diffusion and
target search problems.
Evaluation of $\kappa$ for a given boundary profile is a difficult
task that requires solving the Laplace equation in a complex shape
domain.  Below we present some analytical results for different types
of complex boundaries.  In Sec. \ref{sec:2D}, we consider boundary
profiles and inhomogeneities that are periodic in $x$ coordinate and
translationally invariant along $z$ coordinate, e.g., periodic stripes
or riblets.  The invariance along $z$ coordinate reduces the original
3D problem to a planar one.  In turn, Sec. \ref{sec:3D} focuses on
absorbing spots or deformations, which are periodic in two directions
along the surface.  Section \ref{sec:constant} presents a
constant-flux approximation, which may offer more accurate results,
even though the derived expressions are usually more sophisticated.
In Sec. \ref{sec:channel}, we discuss the Fick-Jacobs equation to deal
with Laplacian transport in channels of variable cross-section and its
application for computing the mean first-passage time (MFPT).  Section
\ref{sec:conclusion} concludes this overview by summarizing the main
steps and discussing other aspects of Laplacian transport.

\section{Periodic patterns in two dimensions}
\label{sec:2D}

\begin{figure}[t!]
\centering
\includegraphics[width=0.3\textwidth]{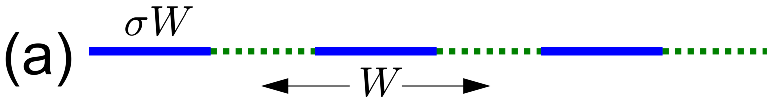}  \hskip 5mm
\includegraphics[width=0.3\textwidth]{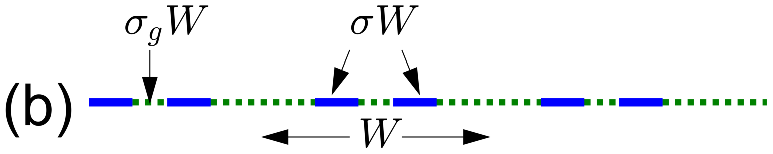}  \hskip 5mm
\includegraphics[width=0.3\textwidth]{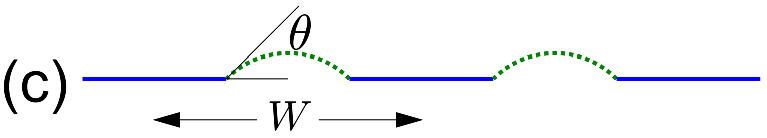}    \vskip 5mm
\includegraphics[width=0.3\textwidth]{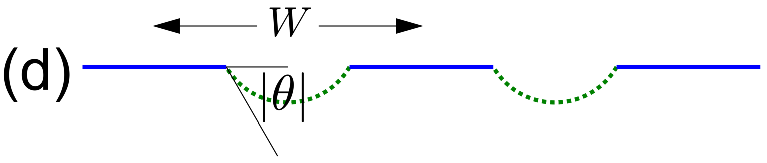}  \hskip 5mm
\includegraphics[width=0.3\textwidth]{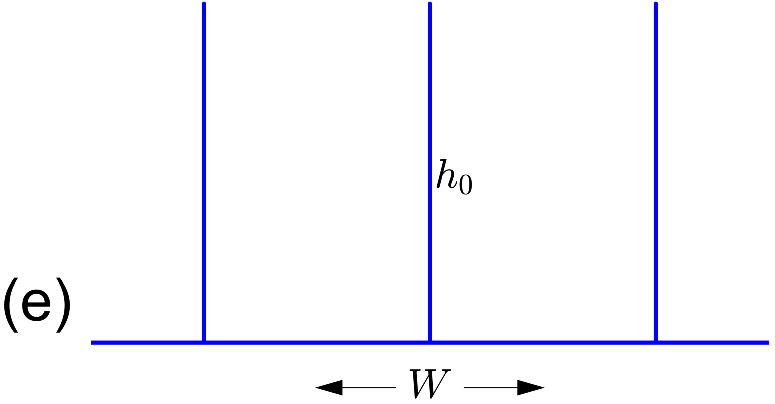}  \hskip 5mm
\includegraphics[width=0.3\textwidth]{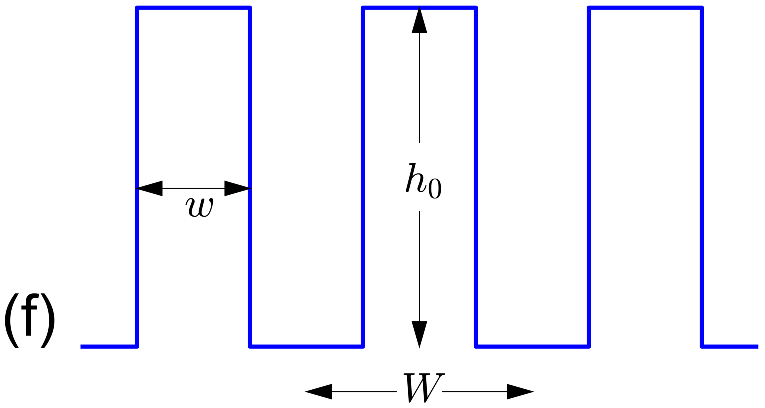}    \vskip 5mm
\includegraphics[width=0.3\textwidth]{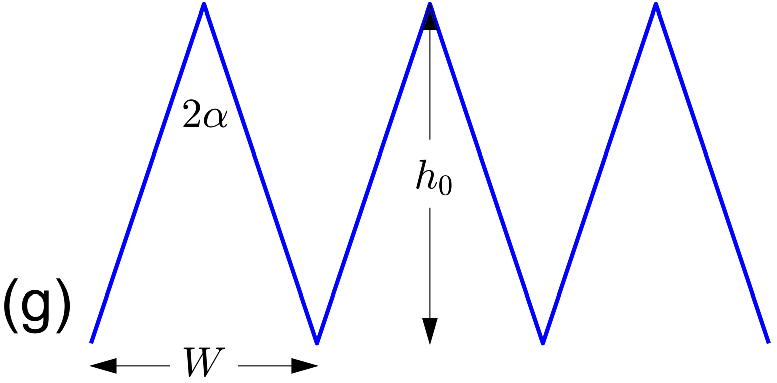}  \hskip 5mm
\includegraphics[width=0.3\textwidth]{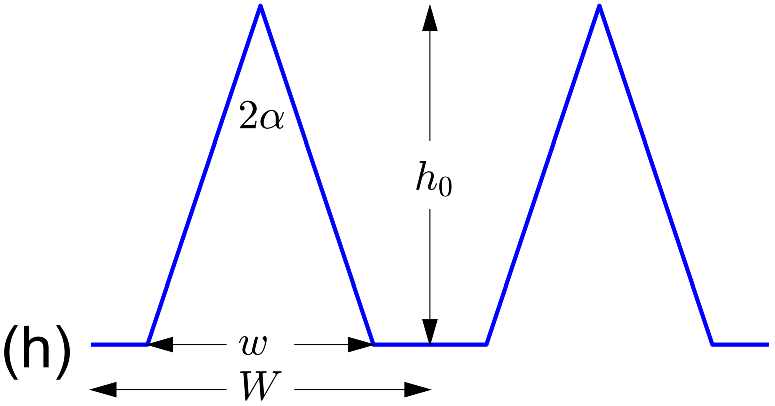}  \hskip 5mm
\includegraphics[width=0.3\textwidth]{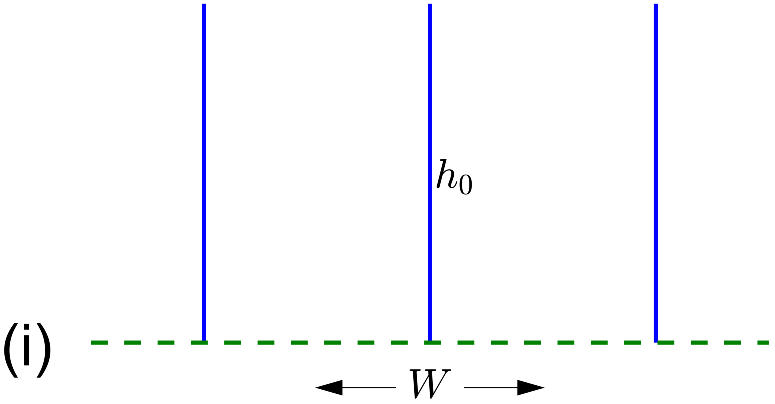}
\caption{ 
Various configurations of absorbing stripes or deformations that are
infinite along $z$ direction, which is perpendicular to the figure.
{\bf (a)} A flat boundary with a periodic arrangement of absorbing
(blue solid) and reflecting (green dashed) stripes, of period $W$ and
absorbing fraction $\sigma$.  {\bf (b)} A flat periodic boundary with
a pattern of two equal absorbing stripes (blue solid) and two unequal
reflecting stripes (green dashed), with the gap $\sigma_g W$ between
two absorbing stripes.  {\bf (c, d)} A periodic boundary with flat
absorbing stripes (blue solid) and non-flat reflecting parts (green
dashed) modeled by circular arcs, with the angle $\theta$ at the joint
point between two stripes: {\bf (c)} upward orientation for $\theta >
0$; {\bf (d)} downward orientation for $\theta < 0$.  {\bf (e)} An
absorbing comb-like boundary.  {\bf (f)} An absorbing boundary with
rectangular riblets of width $w$ and height $h_0$.  {\bf (g)} An
absorbing boundary of saw-tooth riblets of width $W$ and height $h_0$.
{\bf (h)} An absorbing boundary of riblets with trapezoidal grooves.
{\bf (i)} An absorbing comb-like boundary with reflecting base (green
dashed).}
\label{fig:profiles}
\end{figure}

In this section, we focus on various arrangements of infinitely long
stripes or deformations so that the system is invariant along one
direction (we call it $z$), and the original three-dimensional problem
is thus reduced to a two-dimensional problem in the plane $xy$.
Several profiles of stripes and deformations are illustrated in
Fig. \ref{fig:profiles}.
For such profiles, there is a wealth of analytical results for
$\lambda$ and $\kappa$ due to the conformal invariance of the Laplace
equation in two dimensions that enables application of conformal
mapping.  The general framework for evaluation $\lambda$ (and hence
$\kappa$) is given in
Refs. \cite{Crowdy_2011,Vandembroucq_1997,Skvortsov_2014,Skvortsov_2019,Blyth_2003,Hewett_2016,Martin_2014}
and references therein.

\subsection*{Flat boundaries}

For the flat boundary with a periodic alternating pattern of absorbing
and reflecting stripes (Fig. \ref{fig:profiles}a), one has
\cite{Philip_1972} (see also \cite{Moizhes_1955})%
\footnote{
This solution was obtained in \cite{Philip_1972} for a periodic cell,
in which a single absorbing stripe is at the center; in other words,
one solves the Laplace equation $\Delta C(x,y) = 0$ in the
half-infinite stripe, $(-W/2,W/2) \times (0,+\infty)$, with $C = 0$
for $|x|< \tfrac12 \sigma W$ and $y = 0$, $\partial C/\partial y = 0$
for $\tfrac12 \sigma W < |x| < \tfrac12 W$ and $y = 0$, $C \to C_0$ as
$y\to \infty$, and the additional reflecting boundary condition
$\partial C/\partial x = 0$ at two vertical boundaries $|x| = W/2$ and
$y > 0$ of the cell.  This choice of the periodic cell ensures the
periodicity of the solution in the whole upper plane. }
\begin{equation}
\label{ER8}
\lambda = W \frac{\ln [\sin(\pi \sigma/2)]}{\pi} < 0,  \qquad
\kappa = \frac{D}{W} \, \frac{\pi}{\ln [1/\sin(\pi \sigma/2 )]},
\end{equation}   
where $W$ is the period of the pattern, $\sigma$ is the fraction of
absorbing stripes, and we set $y_0 = 0$ to introduce $\kappa$.  As
$\sigma \rightarrow 0$, the trapping parameter $\kappa$ vanishes
logarithmically, $\kappa \propto 1/\ln(1/\sigma) \rightarrow 0$, so
the heterogeneously absorbing boundary becomes perfectly reflecting.
In the opposite limit $\sigma \rightarrow 1$, $\kappa$ diverges as a
power-law, $\kappa \propto 1/(1-\sigma)^2 \rightarrow \infty$, and the
boundary becomes perfectly absorbing.

\begin{figure}[t!]
\centering
\includegraphics[width=0.17\textwidth]{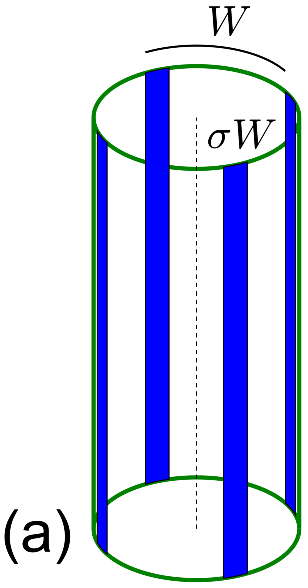}   \hskip 3mm
\includegraphics[width=0.20\textwidth]{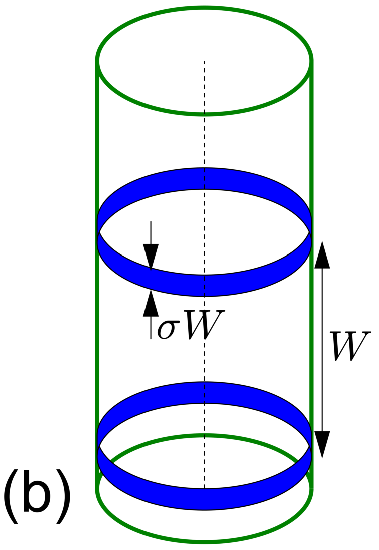}  
\includegraphics[width=0.17\textwidth]{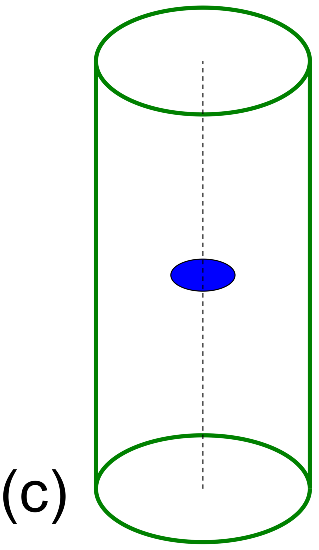}  \hskip 2mm
\includegraphics[width=0.17\textwidth]{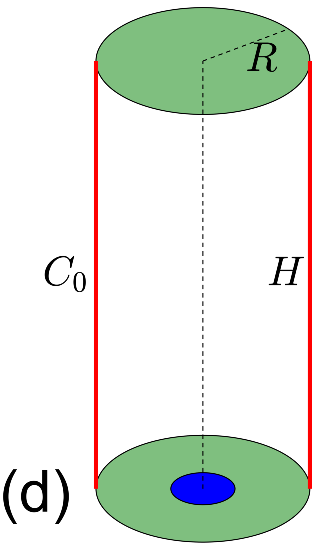}  \hskip 2mm
\includegraphics[width=0.17\textwidth]{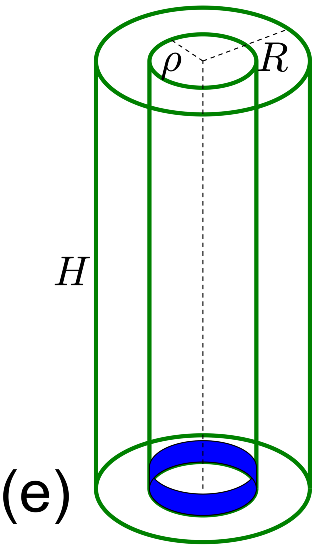}  \vskip 3mm 
\includegraphics[width=0.17\textwidth]{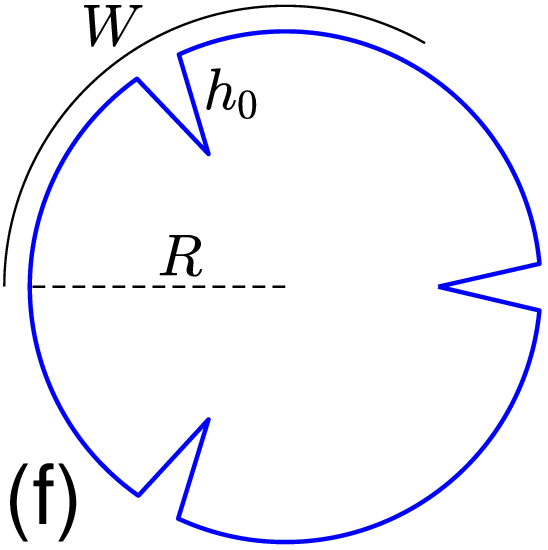} \hskip 3mm
\includegraphics[width=0.17\textwidth]{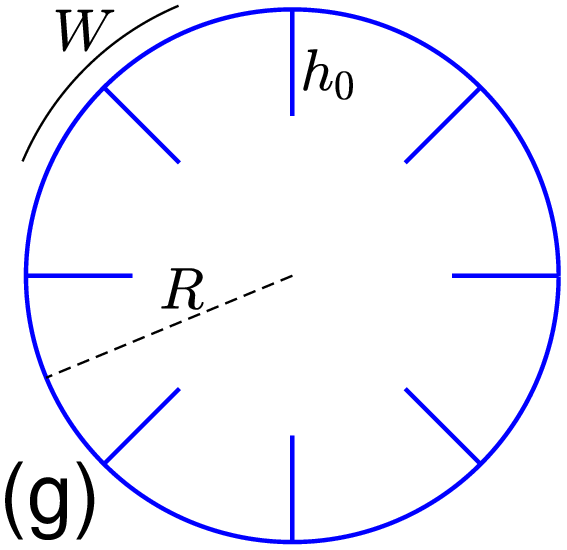} \hskip 3mm
\includegraphics[width=0.17\textwidth]{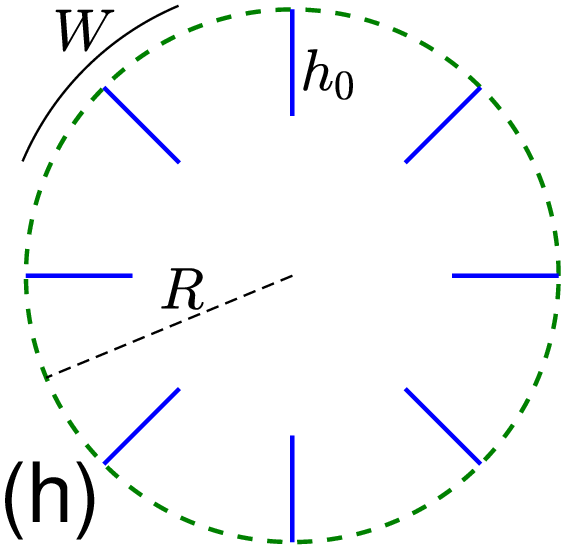} \hskip 3mm
\includegraphics[width=0.33\textwidth]{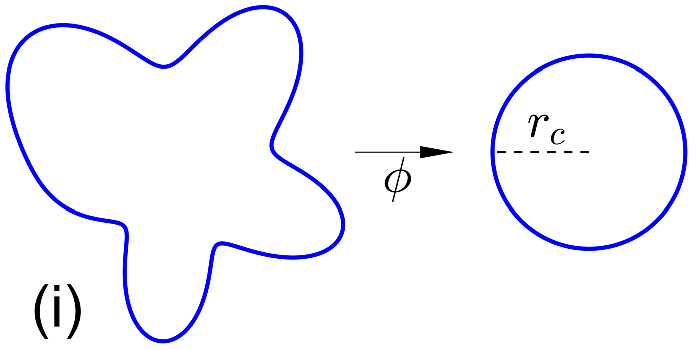} \hskip 3mm
\caption{ 
Cylindrical domains.  {\bf (a)} An infinite reflecting cylinder with
absorbing stripes (in blue) oriented parallel to the cylinder axis.
{\bf (b)} An infinite reflecting cylinder with absorbing stripes (in
blue) oriented perpendicular to the cylinder axis.  {\bf (c)} A long
reflecting cylinder with an absorbing disk (in blue) centered on the
cylinder axis.  {\bf (d)} A capped cylinder of height $H$ and radius
$R$, with the source located at the cylinder wall and an absorbing
disk (in blue), located at the center of the bottom reflecting wall.
{\bf (e)} Two capped coaxial reflecting cylinders of height $H$ and
radii $\rho$ and $R$, with an absorbing stripes (in blue) located on
the inner cylinder.  {\bf (f, g)} Cross-section of an infinite
absorbing cylinder with triangular riblets (the cylinder axis is
perpendicular to the figure).  {\bf (h)} Cross-section of an infinite
reflecting cylinder with absorbing spikes.  {\bf (i)} An appropriate
conformal map $\phi$ can transform an arbitrary simply-connected
cross-section of an infinite cylinder into an equivalent circular
cross-section with a conformal radius $r_c$.}
\label{fig:cylinders}
\end{figure}

It is worth noting that the same result is applicable for
pressure-driven incompressible Stokes flow in a cylindrical pipe with
alternating no-slip and no-stress stripes \cite{Lauga_2003}: when the
stripes are parallel to the cylinder axis (Fig. \ref{fig:cylinders}a),
Eq. (\ref{ER8}) determines the trapping parameter $\kappa$ of the
homogenized cylinder; in turn, if the stripes are perpendicular to the
cylinder axis (Fig. \ref{fig:cylinders}b), Eq. (\ref{ER8}) should be
divided by factor $2$, see details in \cite{Lauga_2003}.  In the
context of target search problems, the trapping parameter $\kappa$ can
be used to estimate the MFPT to the absorbing stripes on the cylinder
from the explicitly known solution for a homogeneous partially
absorbing cylinder:
\begin{equation}  \label{eq:T2r}
T(r) = \frac{R^2 - r^2}{4D} + \frac{R}{2\kappa} ,
\end{equation}
where $r$ is the radial coordinate of the starting point.  In the case
of heterogeneously absorbing cylinder, this is an approximate
solution, which is valid far from the boundary, $R - r \gg W$, where
$W$ is the period.  Moreover, if the starting point is uniformed
distributed in the cylinder, the average of $T(r)$ over the volume
yields
\begin{equation} \label{eq:T2average}
\overline{T} = \frac{1}{\pi R^2} \int\limits_0^{2\pi} d\theta \int\limits_0^R dr \, r \, T(r) = \frac{R^2}{8D} + \frac{R}{2\kappa} \,.
\end{equation} 
Exceptionally, the exact expression for the MFPT in the cylinder with
absorbing parallel stripes is known
\cite{Singer_2006b,Marshall_2016,Grebenkov_2016}, but it is rather
complicated, especially for many stripes.  In turn,
Eqs. (\ref{eq:T2r}, \ref{eq:T2average}), which we use to illustrate an
application of boundary homogenization, provide a rapid and simple way
to estimating the MFPT.

For the configurations of stripes with two and more absorbing stripes
per period, $\kappa$ cannot be described exclusively by the surface
fraction $\sigma$ since the relative position of the absorbing stripes
is also important due to their competition for particles (the
phenomenon known as diffusional screening or diffusive interaction).
The general formula for $\kappa$ is rather cumbersome
\cite{Skvortsov_2020}, but for two equal absorbing stripes per period
(not equally spaced), the result has a simple explicit form:
\begin{equation}  
\label{ER9}
\lambda = W \frac{\ln(F)}{2\pi} < 0,
\qquad F = \sin^2 [\pi (\sigma + \sigma_g )/2] - \sin^2[(\pi \sigma_g )/2],
\end{equation}
where $\sigma$ is the surface fraction of the absorbing stripes, and
$\sigma_g$ is the surface fraction of the gap between them
(Fig. \ref{fig:profiles}b).  For $\sigma_g =0$ we return to the
previous formula (\ref{ER8}), while in the general case $\sigma_g \neq
0$, the parameter $\sigma_g$ describes the diffusive interaction
between the absorbing stripes.  We note that the functional form of
this interaction is nontrivial: even for $\pi (\sigma + \sigma_g )/2
\ll 1$, one has $\lambda \approx W
\ln(\tfrac{\pi^2}{2} \sigma(\sigma_g + \tfrac{1}{2} \sigma))/\pi$,
i.e. $\lambda$ (and thus $\kappa$) is determined by the product
$\sigma (\sigma_g + \tfrac12 \sigma)$, which is a signature of strong
coupling (two tiny stripes strongly interact even when they are
separated by a large distance).
Equations (\ref{ER8}) and (\ref{ER9}) have been extensively validated
numerically with Brownian dynamics simulations, including
homogenization of cylindrical surface with different orientations of
stripes (longitudinal, transverse and spiral)
\cite{Dagdug_2015,Grebenkov_2019a}.

\subsection*{Non-flat boundaries}

Next we consider some examples of non-flat boundaries.  We begin with
the profile, in which the reflecting parts of the boundary are
non-flat.  This boundary has a different value of the parameter
$\kappa$ because the curved reflecting parts affect the particle
ability to reach the absorbing parts of the boundary.  To appreciate
this effect, we can use the results of Ref. \cite{Crowdy_2010}, in
which the reflecting parts were modeled as circular arcs (see
Fig. \ref{fig:profiles}c,d).  In this case, the formula for $\kappa$
includes a correction factor
\begin{equation}
\label{ER11}
\frac{\kappa(\theta)}{\kappa(0)} =  \frac{3 (\pi - \theta)^2}{3 \pi^2 - 4\pi \theta + 2 \theta^2},
\end{equation}   
where $\theta$ is the angle between reflecting and absorbing parts at
their joint, and $\kappa(0)$ is given by Eq. (\ref{ER8}), in which
$\sigma \ll 1$ (dilute limit) was assumed.  At $\theta = 0$, one
retrieves the flat boundary, while negative angles correspond to the
opposite (downwards) orientation of the reflecting arcs.  When
$\theta$ varies from $-\pi/2$ (half-circle oriented downwards) to
$\pi/2$ (half-circle oriented upwards), the correction factor changes
from $1.2273$ to $0.5$, i.e., the boundary becomes less absorbing due
obstruction of the absorbing intervals by the reflecting arcs of the
boundary.  Curiously, the correction factor is larger than $1$ for
$\theta < 0$, i.e., the inclusion of reflecting arcs oriented
downwards enhances the overall reactivity.  Indeed, the particle can
spend more time in these regions and thus have higher chances to
arrive to the absorbing intervals, before returning to the source.
Other examples and their correction factors can be found in
\cite{Crowdy_2016}.  

An important analytical result for the offset $\lambda$ was obtained
for a sinusoidal absorbing boundary of a small amplitude $h_0$
(Fig. \ref{fig:sine}a).  In this case, the formula for $\lambda$ can
be derived by means of perturbation theory \cite{Vandembroucq_1997}:
\begin{equation}
\label{ER12}
\lambda \approx \frac{\pi h^2_0}{W} > 0 , ~~~h_0 \ll W . 
\end{equation}
This formula states that the sinusoidal deformation shifts the
equivalent boundary upwards, though this shift is small since
$\lambda/h_0 \approx \pi h_0/W \ll 1$.  The direction of this shift is
also understandable: the bottom of the troughs of the deformed
boundary is less accessible to the particles, so that most of the
particle absorption occurs near the crests (i.g., above the middle
line located at $y = 0$), which play the role of ``active zones'' of
absorption (see \cite{Sapoval_1999,Filoche_2008} for this concept).
As we can see below, for a small-amplitude deformation of the
boundary, the scaling $\lambda \propto (h^2_0/W)$ is rather general:
it can be interpreted as the product of the amplitude $h_0$ and a
typical slope $\propto h_0/W$ \cite{Vandembroucq_1997}.  We stress
that this result is perturbative; for instance, substitution of $h_0 =
0.5$ and $W = 2$ from the example in Fig. \ref{fig:sine} into
Eq. (\ref{ER12}) yields $\lambda \approx 0.39$, whereas a numerical
solution of the Laplace equation gave $\lambda \approx 0.26$.
Vandembroucq and Roux proposed an improved expression
\begin{equation}
\lambda \approx \frac{2h_0}{\pi} \atan\bigl(\pi^2 h_0/(2W)\bigr),
\end{equation}
which is reduced to Eq. (\ref{ER12}) when $\pi^2 h_0/(2W)$ is small.
For our example, this expression gives $\lambda \approx 0.28$, which
is much closer to the above numerical value.  Note that if the
amplitude $h_0$ was reduced to $0.1$, Eq. (\ref{ER12}) would yield
$0.016$, which is already close to the numerical solution $0.013$.
Expectedly, the perturbative expression (\ref{ER12}) becomes more
accurate as $h_0/W$ gets smaller.

For the comb-like absorbing boundary (Fig. \ref{fig:profiles}e), there
is the exact formula \cite{Skvortsov_2014}
\begin{equation}
\label{ER13}
\lambda = W \, \frac{\ln [\cosh (\pi h_0/W) ]}{\pi} > 0,
\end{equation}
where $h_0$ is the height of the riblets.  This offset is always
positive, i.e., the effective absorbing boundary stands above the base
of the comb.  This is expected because the spikes efficiently absorb
particles and thus increases the diffusive flux $J$, as compared to
the flux $J_0$ onto the absorbing base without spikes.
For $\pi h_0/W \ll 1$, Eq. (\ref{ER13}) leads to a similar scaling as
in Eq. (\ref{ER12})
\begin{equation}
\label{IE00X2azzd}
\lambda \approx \frac{\pi h_0^2}{2W}, ~~~h_0 \ll W.
\end{equation}
The difference by factor $2$ between these expressions can be
attributed to distinct shapes of the sinusoidal profile and the comb
structure (note also that $h_0$ was the amplitude of the sine profile,
not the height).
In contrast, for high riblets $h_0/W \gg 1$, there is a saturation
limit
\begin{equation}
\label{ER14}
\lambda \approx h_0 - W \frac{\ln 2}{\pi}, ~~~h_0 \gg W.
\end{equation}
This result implies that most of the particles are absorbed within the
distance $\frac{\ln 2}{\pi} W \approx 0.22 \, W$ from the tips of the
comb-like structure making the rest of the structure irrelevant in
this context.  This result is the simplest way of illustrating the
concept of ``active zones'' discussed above.  As $\lambda$ remains
below $h_0$, it is natural to impose the effective partially absorbing
boundary at $y_0 = h_0$ (at the tip of the combs), in which case
Eq. (\ref{IE004}) defines the trapping parameter as
\begin{equation}
\kappa = \frac{D}{h_0 - W \ln[\cosh(\pi h_0/W)]/\pi} \,.
\end{equation}
In two limits $h_0/W \ll 1$ and $h_0/W \gg 1$, one gets $\kappa/D
\approx 1/h_0 + \pi/(2W)$ and $\kappa/D \approx (\pi/\ln 2)/W$, 
respectively.

The direct generalization of the comb-like boundary is a periodic
structure with the riblets of a finite thickness.  The closed-form
solution is only available for high rectangular spikes of height $h_0$
and width $w$ (Fig. \ref{fig:profiles}f).  If $h_0/W > 0.6$, one gets
\cite{Richardson_1971}
\begin{equation} 
\label{ER15}
\lambda \approx h_0 - \frac{W}{2 \pi} \mathcal{G}(w/W),
\end{equation}
where $\mathcal{G}(\zeta)$ is the dimensionless function:
\begin{align}
\label{IE00X10}
\mathcal{G}(\zeta) & = 2(1-\zeta) \ln(2-\zeta) + \zeta \ln(2\zeta-\zeta^2), \qquad  \zeta = w/W \le 1.
\end{align}   
In the limit $w\to 0$, one has $\mathcal{G}(0) = 2\ln 2$, so that
Eq. (\ref{ER14}) for the comb-like boundary is retrieved.  In the
opposite limit $\zeta = w/W \to 1$, one has $\mathcal{G}(\zeta)
\approx (1-\zeta)^2$, i.e., the second term in Eq. (\ref{ER15})
vanishes, and one gets $\lambda = h_0$ as expected for a flat
absorbing boundary at the height $h_0$.  As previously, it is natural
to impose the effective partially absorbing boundary at $y_0 = h_0$
(at the tips of the riblets), in which case the trapping parameter is
\begin{equation}
\kappa \approx \frac{D}{W} \, \frac{2\pi}{\mathcal{G}(w/W)} \,.
\end{equation}

The last family of the 2D profiles that we considered in this section
is the saw-tooth boundary (Fig. \ref{fig:profiles}g).  The profile is
characterized by its height $h_0$ and the base width $W$, or by the
saw-tooth angle $2\alpha = 2 \atan(\tfrac12 W/h_0)$.  In this case,
one gets \cite{Bechert_1989}
\begin{equation}
\label{ER17}
\lambda = h_0 - \frac{W}{2 \pi} \mathcal{S}(\alpha),
\end{equation}
where $\mathcal{S}(\alpha)$ is a dimensionless function of angle
$\alpha$: 
\begin{equation}
\label{ER18}
\mathcal{S}(\alpha) = \gamma + 2 \ln 2 + \frac{\pi}{\tan \alpha} - \frac{\pi}{ \alpha} + \psi \left( 1+ \frac{\alpha}{\pi}\right) ,
\end{equation}  
where $\psi (\cdot)$ is the the digamma function, and $\gamma \approx
0.5772$ is the Euler constant.
One can check that $\mathcal{S}(\alpha)$ is a monotonously decreasing
function of $\alpha$.  For $\alpha =0$ (comb-like boundary), one has
$\psi (1) = - \gamma$ and thus $\mathcal{S}(0) =2 \ln 2$, in agreement
with Eq. (\ref{ER14}).  For $\alpha =\pi/2$ (flat surface), one gets
$\psi (3/2) = - \gamma - 2 \ln 2 + 2$, so that $\mathcal{S}(\pi/2)
=0$, as expected.  Another explicit result corresponds to $\alpha =
\pi/6$ (equilateral triangle), for which $\mathcal{S}(\pi/6) =
\pi\sqrt{3}/2 - 3\ln(3)/2$ and thus $\lambda/W = \sqrt{3}/4 + 3 \ln
(3)/(4 \pi)$ \cite{Blyth_2003}.  As previously, setting $y_0 = h_0$,
one can also introduce the trapping parameter as
\begin{equation}
\kappa = \frac{D}{W} \, \frac{2\pi}{\mathcal{S}(\alpha)} .
\end{equation}

More complex profiles appear when the saw-teeth are separated by a
piece of flat boundary, with a period $W$ (the so-called trapezoidal
grooves, Fig. \ref{fig:profiles}h).  The generalized formulas were
proposed in Ref. \cite{Bechert_1989}, e.g., for high riblets, one has
\begin{equation}
\label{ER19}
\lambda \approx h_0 - \biggl[(W - w) \frac{\ln 2}{\pi}  + \frac{w}{2 \pi} \mathcal{S}(\alpha)\biggr], ~~~ h_0/W > 1,
\end{equation}   
where $w$ is the width of the foot of the riblet, and $\alpha = \atan
(\tfrac12 w/h_0)$.  Moreover, one can account for rounded wedges at
the tips \cite{Bechert_1989}.
More numerical and analytical estimates of $\lambda$ for different
types of boundaries (piece-wise, pre-fractals, convex, with scalloped
and club-like riblets, etc.) can be found in Ref. \cite{Bechert_1989}.
We also mention an interesting result established numerically in Ref.
\cite{Blyth_2003} regarding the strong non-monotonic dependence of the
parameter $\lambda$ on geometrical parameters of $M$-shape profiles.

To conclude this section, we present an example of non-flat boundaries
with mixed boundary conditions (absorbing and reflecting).  For the
comb-like boundary with absorbing spikes and the reflecting base
(Fig. \ref{fig:profiles}i), the result is \cite{Skvortsov_2019}
\begin{equation}  
\label{ER22}
\lambda = W\, \frac{\ln [\sinh (\pi h_0/W) ]}{\pi} .
\end{equation}  
For $\pi h_0/W \gg 1$, the saturation limit, $\lambda \approx h_0 - W
\ln(2)/\pi$, remains the same as for the absorbing base (since
particles do not reach the bottom of the comb structure, the boundary
condition at the base becomes irrelevant).  As the comb height $h_0$
decreases, the offset parameter $\lambda$ also decreases and then
becomes negative.  In the limit $h_0/W \ll 1$, the offset parameter
diverges to $-\infty$ logarithmically, and the boundary slowly becomes
perfectly reflecting.  In this limit, Eqs. (\ref{ER8}) and
(\ref{ER22}) become identical provided we set $h_0/W = \sigma/2$.  The
factor $2$ can be explained by symmetry reasoning: a horizontal
absorbing interval laying on the reflecting base creates the diffusive
flux twice smaller than the flux created by the same vertical interval
standing on that base (due to two accessible sides of this interval).
These results can be extended to incorporate other shapes of small
absorbers by finding an equivalent absorbing interval with the same
log-capacity and using it in Eq. (\ref{ER8}).  An effective radiative
boundary can be set at $y_0 = h_0$ with the trapping parameter
\begin{equation}
\kappa = \frac{D}{h_0 - W\, \ln [\sinh (\pi h_0/W) ]/\pi} ,
\end{equation}
which vanishes very slowly in the limit $h_0 \to 0$.

As earlier in the case of cylinders with absorbing stripes, the above
expressions for the trapping parameter $\kappa$, which were derived
for a flat boundary, are applicable to curved surfaces.  For instance,
if the cylinder of radius $R$ contains grooves or riblets that are
parallel to the cylinder axis (Fig. \ref{fig:cylinders}f,g,h), one can
get $\kappa$ by choosing an appropriate expression according to the
groove shape, and then substitute it to Eqs. (\ref{eq:T2r},
\ref{eq:T2average}) for estimating the MFPT.  Even though the height
$h_0$ of these grooves and the period $W$ are formally expected to be
relatively small with respect to $R$ (which plays here the role of $H$
for the flat boundary), such an estimate remains quite accurate for a
broad range of parameters.  Moreover, there is no smoothness
constraint on such a cylindrical surface with grooves or riblets
(e.g., one can even consider a comb-like structure).  This feature
reveals the crucial role of conformal mapping, on which many of the
above expressions rely, and thus distinguishes them from other
approximations obtained, e.g., via perturbation theory.  More
generally, for a cylinder of any simply-connected cross-section
$\Omega$ (Fig. \ref{fig:cylinders}i), one can construct a conformal
map from $\Omega$ onto a disk and thus determine its conformal radius
$r_c$, from which the MFPT can be estimated by replacing $R$ by $r_c$
in Eq. (\ref{eq:T2average}) and setting $\kappa = \infty$.  It is
important to stress the two-step character of this approximation: one
first estimates $r_c$ and then evaluates the MFPT for an equivalent
disk.  In contrast, even though a direct application of conformal
mapping to the first-passage problem is possible (see
\cite{Grebenkov_2016} for details), the Poisson equation governing the
MFPT is not conformally invariant that brings substantial
complications to the exact solution.
Despite the simplicity and potential utility of boundary
homogenization, its applications are not yet fully explored in the
context of target search problems, while systematic investigations on
their accuracy and validity range are still missing.

\section{Periodic patterns in three dimensions}
\label{sec:3D}

When the domain is not translationally invariant along one axis, the
original 3D problem is not reducible to a planar one that limits the
use of conformal mapping.  As a consequence, the analysis of patterns
that are periodic in two directions, is much more difficult, with a
limited number of available analytical results.

\subsection*{Flat boundaries}

Similar to the 2D case, we begin with the trapping parameter of a
smooth boundary covered by absorbing and reflecting spots.  The first
results by Berg and Purcell \cite{Berg_1977,Berg_1993} and by Zwanzig
\cite{Zwanzig_1990} dealt with the particle trapping by small
absorbing spots covering a reflecting sphere.  The opposite case (a
small reflecting spot on the absorbing sphere) was analyzed in
\cite{Dagdug_2016}.  This is still an area of active research
\cite{Lindsay_2017,Handy_2021}.

\begin{figure}[t!]
\centering
\includegraphics[width=0.44\textwidth]{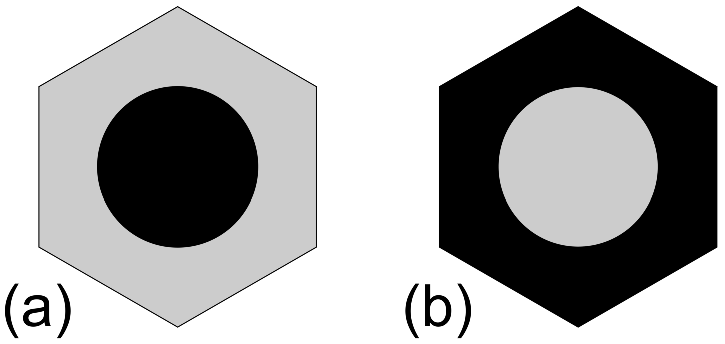}  \hskip 5mm
\includegraphics[width=0.44\textwidth]{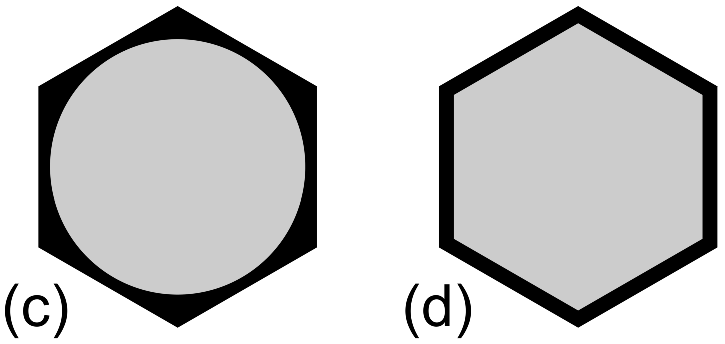} 
\caption{
Periodic cells on a hexagonal lattice with mixed absorbing-reflecting
properties (dark area is absorbing and light area is reflecting): {\bf
(a)} absorbing disk on the reflecting base; {\bf (b)} reflecting disk
on the absorbing base; {\bf (c)} a large reflecting disk on the
absorbing base, which is equivalent to a thin polygonal absorbing ring
on the reflecting base {\bf (d)}.}
\label{fig:hexagon}
\end{figure}

The appeal to electrostatic analogy provides a valuable insight and
clear interpretation of the main result on this topic
\cite{Krapivsky_2010,Berg_1993}.  According to this analogy, the trapping
capacity $K$ of an object and its electric capacity (or capacitance)
$\mathcal{C}$ are related as $K = \mathcal{C} D$
\cite{Krapivsky_2010,Berg_1993} (here we employ the convention that
the capacity of a sphere of radius $R$ is $4\pi R$).  There are many
analytical methods for estimating $\mathcal{C}$ from the shape of the
object \cite{Smythe_1972,Chow_1982,Dudko_2004,Berezhkovskii_2007b}
that enables fruitful translation of many results from electrostatic
to diffusion kinetics and target search problems.

We begin with an illustrative example: a sparse periodic array of
absorbing disks of radius $a$ on a reflecting surface
(Fig. \ref{fig:hexagon}a).  If diffusional screening between disks
could be ignored, an estimation of the trapping parameter for such a
boundary would be straightforward \cite{Hill_1975}
\begin{equation}
\label{ER27}
\kappa = \frac{K}{\mathcal{A}_0} = \frac{\mathcal{C} D} {\mathcal{A}_0} = \frac{4 a D} {\mathcal{A}_0}, 
\end{equation}
where $\mathcal{A}_0$ is the area of the periodic cell, and
$\mathcal{C} =4 a$ is half of the capacity of a disk in the space
$\R^3$ (the factor $1/2$ accounts for the half of the flux to one side
of the disk surface).  Even though diffusional screening between disks
is always present, the formula (\ref{ER27}) is still valid if
$\mathcal{C}$ is understood as the capacity of a disk in the lattice,
which depends on the arrangement of the disks.  In the formal limit
when the surface fraction of absorbing disks $\sigma$ is equal to $1$
(full coverage), one retrieves the infinite capacity of the perfectly
conducting plane: $\mathcal{C} = \infty$.  The calculation of the
``renormalized'' $\mathcal{C}$ (and hence $\kappa$) that accounts for
the diffusive interaction between absorbers in the lattice is the main
aim of the effective medium approximation in this context.  Different
analytical methods provide different levels of rigor and accuracy.

The aforementioned electrostatic analogy allows us to reveal some
important features of boundary homogenization in three dimensions.
The trapping parameter $\kappa$ as function of the surface fraction
$\sigma$ of the absorbing part can be strongly dependent on its shape.
To illustrate this point, we analyze two complementary configurations:
a ``spot configuration'' (Fig. \ref{fig:hexagon}a) formed by a
hexagonal lattice of absorbing disks ($C =0$) on the reflecting base
($dC/dy =0$), and a ``mesh configuration'' (Fig. \ref{fig:hexagon}b),
which is an inverse of the first one: the same lattice but of the
reflecting spots ($dC/dy =0$) on the absorbing base ($C =0$).  In the
limit of small $\sigma$, the first configuration reduces to a sparse
periodic arrangement of small disks, so according to Eq. (\ref{ER27}),
one has $\kappa \propto a \propto \sqrt{\sigma}$, while the second
configuration transforms to a lattice of flat rings
(Fig. \ref{fig:hexagon}c,d) and $\kappa$ decays much slower: $\kappa
\propto 1/\ln(1/\sigma)$ (this follows from the formula for the
capacity of a flat ring \cite{Leppington_1972}, see also
Eq. (\ref{ER8}) for the stripe configuration).  This comparison
illustrates that the knowledge of $\sigma$ alone is insufficient for
estimating the trapping parameter without additional information on
geometrical settings.

It is instructive to pursue the analysis of these two configurations
and to deduce explicit formulas for $\kappa$ and then by comparison to
appreciate their differences.  Our choice of the hexagonal lattice and
of the circular shape of absorbing/reflecting spots helps to avoid
additional complications related to anisotropy or other asymmetries.
To simplify analytical treatment, the periodic cell of this system is
often approximated by a circular tube of the same cross-section
containing a single disk \cite{Keller_1967}.

\subsubsection*{Absorbing disk on reflecting base}

For the first configuration (Fig. \ref{fig:hexagon}a), it is
convenient to write the trapping parameter $\kappa$ in the following
form \cite{Berezhkovskii_2004,Berezhkovskii_2006,Muratov_2008}
\begin{equation}
\label{ER28}
\kappa  = \frac{D}{\sqrt{\mathcal{A}_0}}  F_1(\sigma) , 
\end{equation}
where $\sigma$ is the surface fraction of absorbing area, and
$\mathcal{A}_0$ is the area of the periodic cell of the lattice.  In
turn, the dimensionless function $F_1(\sigma)$ can be written as
$F_1(\sigma) = \frac{4}{\sqrt{\pi}} \sqrt{\sigma} f(\sigma)$ to ensure
$f(0) = 1$ according to Eq. (\ref{ER27}) in the limit $\sigma\to 0$.
The asymptotic behavior of the function $f(\sigma)$ as $\sigma
\rightarrow 0$ and $\sigma \rightarrow 1$ can be deduced from the
known analytical results, including the capacitance approximation.

In 1941, Fock studied the steady flow of electric current down a long
circular tube filled with a conducting medium, with a thin circular
hole in its center (Fig. \ref{fig:cylinders}c).  In particular, Fock
obtained a representation of the function $f(\sigma)$ (and hence
$F_1(\sigma)$) as a power-law series of $\sqrt{\sigma}$, evaluated the
coefficients of this series in terms of the integrals of Bessel
functions, and gave numerical values of the first 12 coefficients
\cite{Fock_1941}.  Fock also proposed a simple truncated expression
\begin{equation}
\label{ER29}
  f(\sigma) \approx \frac{1}{1 - 1.41\sigma^{1/2} + 0.34\sigma^{3/2} + 0.07 \sigma^{5/2}} .
\end{equation}
This expression provides a very accurate behavior of $f(\sigma)$ as
$\sigma \to 0$.  Moreover, it ensures the expected divergence of
$f(\sigma)$ as $\sigma \to 1$ but fails to yield the correct power-law
exponent.  This issue was resolved by Leppington and Levine
\cite{Leppington_1972}, who derived the asymptotic behavior near
$\sigma \rightarrow 1$:
\begin{equation} 
\label{IE12XxFFF}
  f(\sigma) \approx  \frac{1}{2(1- \sqrt{\sigma})^2} \approx \frac{2}{(1- \sigma)^2} , ~~~ \sigma \rightarrow 1,
\end{equation}
where the second relation was obtained by writing $1-\sigma =
(1-\sqrt{\sigma})(1+\sqrt{\sigma}) \approx 2(1-\sqrt{\sigma})$ as
$\sigma\to 1$.  Motivated by the relative smallness of the coefficient
of the last term in the denominator of Eq. (\ref{ER29}) and combining
two asymptotic results together, one arrives at the following approximation
\begin{equation}
\label{ER30}
 F_1(\sigma) \approx \frac{4}{\sqrt{\pi}} \sqrt{\sigma} \, 
\underbrace{\frac{1 + A\sqrt{\sigma} - (A-1)\sigma^{3/2}}{(1 - \sigma)^2}}_{\approx f(\sigma)} ,
\end{equation}
which is applicable for any $0 < \sigma < 1$.  The coefficient $A =
1.41$ is fixed by matching the coefficient in front of $\sqrt{\sigma}$
in the Taylor expansion of the Fock's expression (\ref{ER29}) as
$\sigma\to 0$, whereas the coefficient $A-1$ in front of
$\sigma^{3/2}$ ensures the correct asymptotic behavior
(\ref{IE12XxFFF}) as $\sigma\to 1$.  We stress that this approximation
has no fitting parameter.  Note that Bernoff {\it et al.} derived a
similar result by other means \cite{Bernoff_2018}.  A very similar
approximation was earlier proposed by Berezhkovskii {\it et al.} by
fitting the numerical solution of the Laplace equation
\cite{Berezhkovskii_2004,Berezhkovskii_2006,Muratov_2008}
\begin{equation} 
\label{ER30_bis}
 F_1(\sigma) \approx \frac{4}{\sqrt{\pi}} \sqrt{\sigma} \, 
\underbrace{\frac{1 + A\sqrt{\sigma} - B\sigma^2}{(1 - \sigma)^2}}_{\approx f(\sigma)} .
\end{equation}
The parameters $A$ and $B$ were estimated from fitting and were shown
to depend on the considered lattice, e.g., $A \approx 1.37$ for the
hexagonal lattice
\cite{Berezhkovskii_2004,Berezhkovskii_2006,Muratov_2008}.  Note that
the second parameter $B$ should be fixed as $A-1$ to ensure the
correct asymptotic behavior (\ref{IE12XxFFF}) as $\sigma \to 1$.

One can compare Eq. (\ref{ER28}) with other results for $\kappa$ found
in the literature \cite{Bernoff_2018}.  The solution of the celebrated
Berg-Purcell problem \cite{Berg_1977} for a reflecting sphere covered
by a large number $\mathcal{N} \gg 1$ of small absorbing disks of
radius $a$ reads in our notations as
\begin{equation} 
\label{ER35}
\kappa_{\rm BP} = \frac{D a \mathcal{N} }{\pi R^2} = \frac{4D}{\pi a} \, \sigma ,
\end{equation}
where $\sigma = \mathcal{N} (a/2 R)^2$.  As the Berg-Purcell solution
ignores diffusional screening between absorbing spots, it is valid
only for $\sigma \ll 1$ (the dilute limit).
Later Zwanzig heuristically modified the Berg-Purcell result
(\ref{ER35}) to explicitly account for the reduction of the surface
area due to absorbing disks \cite{Zwanzig_1990}:
\begin{equation}
\label{IE12}
\kappa_{\rm ZW} =  \frac{\kappa_{\rm BP}}{1 - \sigma} .
\end{equation}
While the diffusive transport towards a spherical boundary is quite
different from that on the plane, one can still compare the
predictions given by Eqs. (\ref{IE12}) and (\ref{ER28}).  In the
latter case, the surface area of the ``periodic cell'' can be written
as $\mathcal{A}_0 = 4\pi R^2/\mathcal{N}$ so that Eq. (\ref{ER28})
becomes
\begin{equation}  \label{eq:kappa_sphere}
\kappa \approx \frac{D}{a \sqrt{\pi}} \, \sqrt{\sigma} F_1(\sigma) = \frac{4D \sigma}{\pi a}\, f(\sigma).
\end{equation}  
As outlined by Bernoff {\it et al.} \cite{Bernoff_2018}, Zwanzig's
correction (\ref{IE12}) yields a slower divergence, $(1 -
\sigma)^{-1}$ instead of $(1-\sigma)^{-2}$, as $\sigma \to 1$.  In
turn, Eq. (\ref{eq:kappa_sphere}), which was originally derived by
homogenizing the flat boundary, is applicable for the whole range of
$\sigma$ even for a sphere.

\subsubsection*{Reflecting disk on absorbing base}

For the second configuration (Fig. \ref{fig:hexagon}b), we also write
\begin{equation}
\label{ER30_tre}
\kappa = \frac{D}{\sqrt{\mathcal{A}_0}} F_2(\sigma) .
\end{equation} 
An important insight comes from the identity $\kappa = D/\mathcal{B}$
\cite{Martin_2022,Martin_2022a,Skvortsov_2023b}, where $\mathcal{B}$
is the blockage coefficient of a potential flow in a tube with a
blocking disk.  The value of the blockage coefficient was calculated
in \cite{Martin_2020,Martin_2022}, yielding the following asymptotic
behaviors
\begin{subequations}
\begin{align}
\label{ER32}
F_2(\sigma) & \approx \frac{{\pi}^{3/2}}{\ln (2/\sigma)} , ~~~  \sigma \ll 1, \\
\label{ER31}
F_2(\sigma) & \approx \frac{3 {\pi}^{3/2}} {4 (1-\sigma)^{3/2}} , ~~~  \sigma \rightarrow 1.
\end{align}
\end{subequations}
Combining two asymptotic results, one arrives at the following
interpolation
\begin{equation} 
\label{ER33}
F_2(\sigma) \approx \frac{3 {\pi}^{3/2}} {4 (1-\sigma)^{3/2} \ln \bigl(P + Q \sigma^{-4/3}\bigr)} , 
\end{equation}
with two constants $P$ and $Q$ that follow by matching
Eq. (\ref{ER33}) with the asymptotic relations (\ref{ER32},
\ref{ER31}):
\begin{equation}
\label{ER34}
P = \exp(1) - 2^{4/3} \approx 0.20 , ~~~  Q = 2^{4/3} \approx 2.51 .
\end{equation}
While the values of constants $A$, $B$, $P$, $Q$ are affected by the
lattice type (see
\cite{Berezhkovskii_2004,Berezhkovskii_2006,Muratov_2008,Bernoff_2018}
for details), the functional form of $F_1$ and $F_2$ remains
unchanged.  As illustrated on Fig. \ref{fig:hexagon}c,d, the limit
$\sigma \to 0$ is particularly interesting because such a
configuration forms a sort of spider web of thin flat absorbing rings
on the hexagonal lattice.  Moreover, using the capacitance argument,
one can replace thin flat stripes by thin absorbing cylinders whose
arrangement resembles a graphene layer.  While a direct computation of
the trapping parameter for such a structure is challenging,
Eq. (\ref{ER33}) can provide a first approximation.

The expression (\ref{ER33}) for the function $F_2(\sigma)$ can be
compared with the results of \cite{Dagdug_2022a,Skvortsov_2023b}
deduced from fitting the numerical solution:
\begin{equation}
\label{ER36}
F_2(\sigma) = \frac{3 {\pi}^{3/2} g(\zeta)}{4 \zeta^3 [1 - \ln(1- \zeta)]} , 
\end{equation}
where $\zeta = \sqrt{1- \sigma}$, $g(\zeta) =1 + 0.6 \zeta + 2
\zeta^2 - 1.5 \zeta^3 - 0.8 \zeta^{100}$, and $\sigma$ is the surface
fraction of absorbing parts.  If Eq. (\ref{ER36}) may be considered as
a benchmark, such a comparison reveals a moderate quality of the
interpolation formula (\ref{ER33}).  Moreover, the asymptotic relation
(\ref{ER32}) is accurate only at very small $\sigma$, whereas
next-order corrections are needed for moderately small $\sigma$.  One
can also compare these results to the capacitance approximation
$\kappa = D \mathcal{C}/\mathcal{A}_0$ with the capacity $\mathcal{C}$
of a flat circular ring, given in \cite{Leppington_1972}.  Further
investigations could clarify these points and bring more accurate
approximations.

\subsubsection*{Hindering effect of a porous membrane}

So far we were focused on the effect of mixed absorbing/reflecting
patterns on the plane at $y=0$.  In a similar way, boundary
homogenization can be employed to estimate the hindering effect of
pores inside a semi-permeable membrane.  Let us consider a flat
absorbing boundary at $y = 0$, which is separated from the source at
height $H$ by a semi-permeable membrane of thickness $w$, located at
some height $h_0$.  The presence of the membrane reduces the diffusive
flux from the basic value $J_0 = D C_0/H$ without membrane, to
\cite{Skvortsov_2023d,Skvortsov_2021c}
\begin{equation}
J = \frac{C_0}{\frac{H-w}{D} + \frac{w}{D\sigma} + \frac{2}{\kappa_m}} ,
\end{equation}
where $\sigma$ is the surface fraction of pores in the membrane, and
$\kappa_m$ is given by Eq. (\ref{ER28}) or Eq. (\ref{ER30}), depending
on the arrangement of pores.  As discussed in Sec. \ref{sec:intro},
the ratio $C_0/J$ can be interpreted as the overall diffusive
resistance (or impedance) of the system, allowing for a simple
physical interpretation: the bulk resistance $(H-w)/D$ of the region
without membrane is sequentially added to the resistance of the
membrane.  The latter has two contributions: the resistance
$w/(D\sigma)$ of multiple pores (connected in parallel) and the
additional contributions $2/\kappa_m$ accounting for heterogeneous
concentration of arrived particles onto the membrane.  Expectedly, the
overall resistance and thus the flux do not depend on the height $h_0$
due to the sequential addition of resistances.  As a consequence,
Eqs. (\ref{eq:lambda_def}, \ref{IE004}) define
\begin{equation}
\lambda = w(1 - 1/\sigma) - \frac{2D}{\kappa_m} < 0 ,  \qquad \kappa = \frac{1}{\frac{w}{D}(1/\sigma - 1) + \frac{2}{\kappa_m}} ,
\end{equation}
where the effective radiative boundary was set at $y_0 = 0$.
Expectedly, the trapping parameter $\kappa$ decreases when the width
$w$ of the membrane increases or the pores vanish ($\sigma\to 0$).
Note that the factor $2$ in the contribution $2/\kappa_m$ comes from
two sides of the membrane; if the membrane lies on the absorbing
boundary, $2/\kappa_m$ should be replaced by $1/\kappa_m$.

\subsection*{Non-flat boundaries}

There is only a limited number of analytical results for the trapping
parameter $\kappa$ for the boundaries with complex 3D profiles (e.g.,
boundaries with protrusions).  Some of these results will be presented
below.

Similar to the 2D sinusoidal profile (Fig. \ref{fig:sine}a), one can
analyze a wavy absorbing surfaces in three dimensions (i.e., a
doubly-periodic sinusoidal deformation with a square or hexagonal
pattern).  The problem can be treated by perturbation theory that
yields \cite{Fyrillas_2001}
\begin{equation}
\label{ER37}
 \lambda \approx q h^{2}_0/W,
\end{equation}
where $q = 2 \pi$ for the square pattern and $q =4 \pi/\sqrt{3}$ for
the hexagonal pattern (note a weak dependence on the lattice
geometry).  These results were validated by numerical simulations
\cite{Fyrillas_2001}.  For small values of $h_0/W$ (up to 0.1), the
agreement between the asymptotic and numerical results is good,
confirming the parabolic scaling near the origin in Eqs. (\ref{ER12},
\ref{ER37}) for both 2D and 3D inhomogeneities.

\begin{figure}[t!]
\centering
\includegraphics[width=0.32\textwidth]{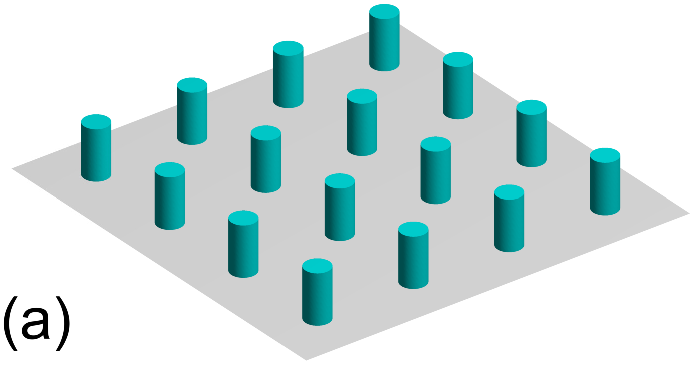} 
\includegraphics[width=0.33\textwidth]{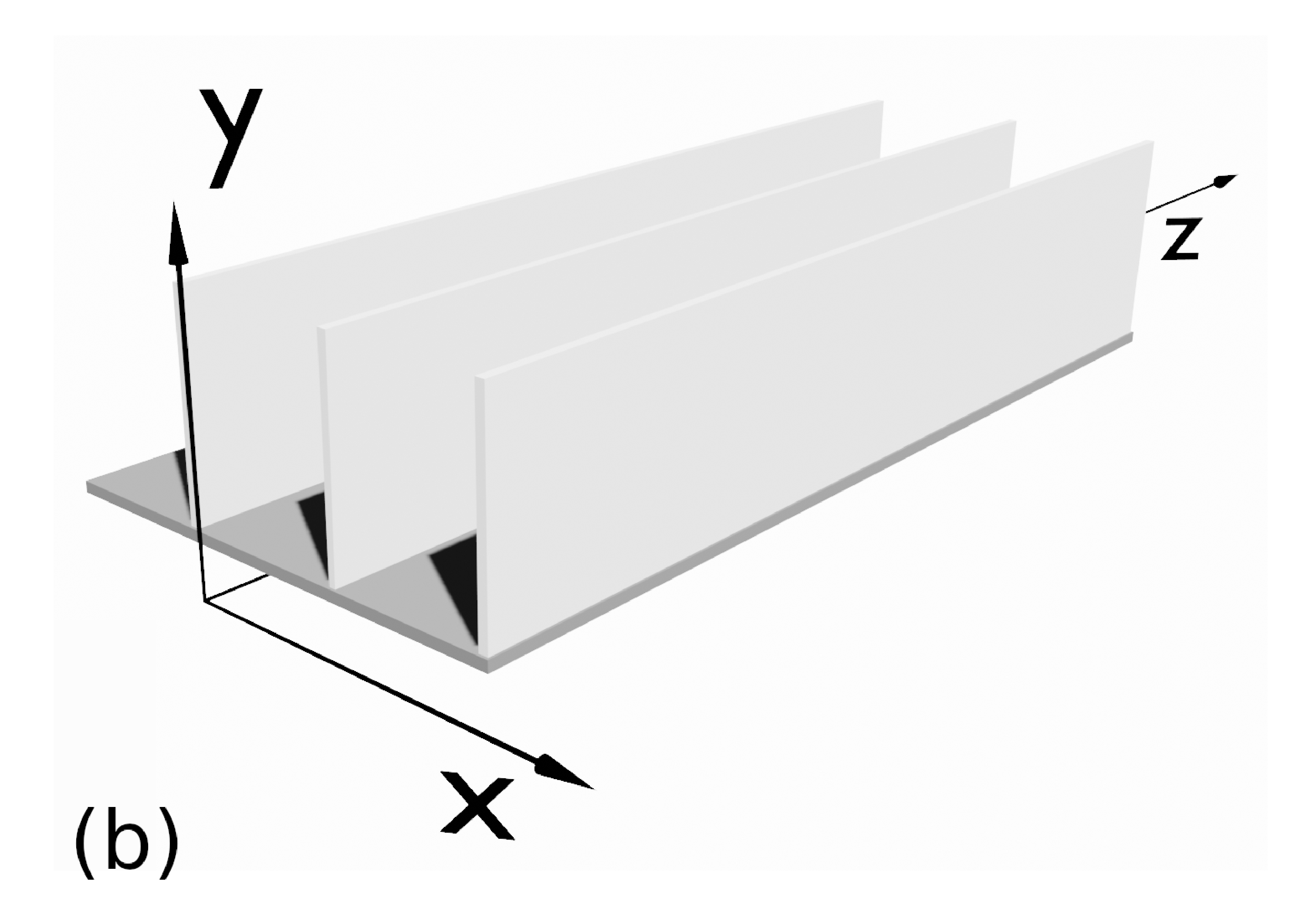} 
\includegraphics[width=0.33\textwidth]{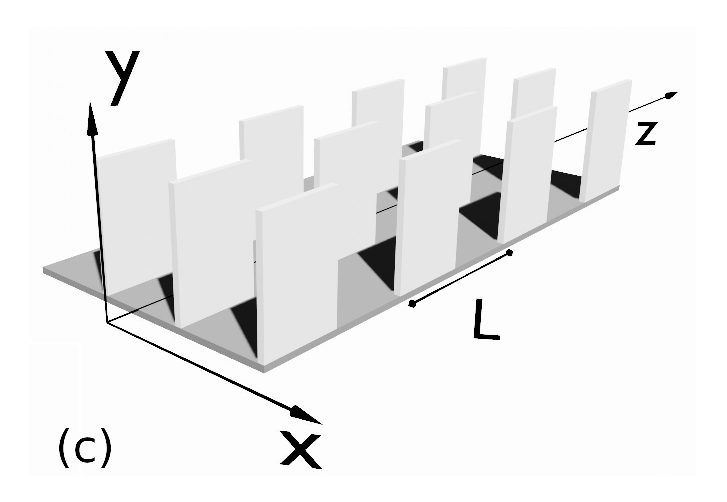} 
\caption{
Examples of irregular boundaries: {\bf (a)} a square lattice of
absorbing pillars on a flat boundary; {\bf (b,c)} a 3D coating of
periodic riblets without or with gaps.}
\label{fig:pillars}
\end{figure}

A lattice of absorbing/reflecting cylinders (pillars) is the natural
extension of the model of rectangular riblets considered in
Sec. \ref{sec:2D}.  For a square lattice of absorbing pillars on the
absorbing base (Fig. \ref{fig:pillars}a), the trapping parameter was
computed and analyzed in \cite{Grebenkov_2022b}.  Since the exact
expression for $\kappa$ is rather cumbersome, we just mention on two
limits.

(i) For thin and tall pillars, one gets
\begin{equation}
\label{ER38}
\lambda \approx h_0 - W  \sqrt{ \ln (1/\sigma) - 3/4}, \quad h_0 \gg W,~ \sigma \ll 1,
\end{equation}
where $W$ is the lattice period and $\sigma$ is the surface fraction
occupied by the tips of the pillars.  Setting an effective radiation
boundary at $y_0 = h_0$, one gets 
\begin{equation}
\kappa \approx \frac{D}{W} \, \frac{1}{\sqrt{ \ln (1/\sigma) - 3/4}} \,.
\end{equation}
Similar to the 2D case, this formula is independent of $h_0$ since the
particles are mostly absorbed near the top of the pillars.  

(ii) For thin and short pillars, one has
\begin{equation}
\label{ER39}
\lambda \approx h_0 -  W^2 \frac{ \ln (1/\sigma) - 3/4}{h_0}, \quad
\kappa \approx \frac{D}{W} \, \frac{h_0/W}{\ln (1/\sigma) - 3/4},  \quad h_0 \ll W,~ \sigma \ll 1.
\end{equation}
In this regime, the value of $\lambda$ is affected by the height
$h_0$: as the pillar gets shorter, $\lambda$ decreases, as expected.
The limit of $h_0 =0$ is given by Eq. (\ref{ER30}).  In both regimes
of high and short pillars, one has $\lambda \rightarrow -\infty$ and
$\kappa \to 0$ in the limit $\sigma \rightarrow 0$ because infinitely
thin pillars (needles) are inaccessible to diffusion in three
dimensions.  This is in sharp contrast to the planar case, for which
rectangular riblets of width $w$ were reduced to a comb-like boundary
as $\sigma = w/W \to 0$, and the latter was still accessible to
diffusing particles; in particular, the offset $\lambda$ and the
trapping parameter $\kappa$ remain finite, see Eq. (\ref{ER22}).

An approximation for the offset $\lambda$ can be derived analytically
for a coating made of a periodic system of absorbing riblets with
periodic gaps in each riblet (Fig. \ref{fig:pillars}b,c).  This seems
to be one of the simplest realizations of a 3D non-differentiable
profile.  The result is \cite{Skvortsov_2023c}
\begin{equation}
\label{ER40}
\lambda \approx \frac{W - 2 \lambda_s}{\pi} \ln ( \cosh [\pi h_0 /(W - 2  \lambda_s) ]),
\end{equation}
where $W$ is the distance between riblets,
\begin{equation}
\label{IE16xxzz}
\lambda_s  = \frac{L}{\pi} \ln [\sin(\pi \sigma_s/2)] < 0,
\end{equation}
$L = s + g$ is the period of the solid-gap structure of an individual
riblet, $s$ is the width of the solid part of the riblet (per period),
$g = L - s$ is the width of the gap, and $\sigma_s = s/L$.  For the
case $g = 0$ (no gaps), one has $\sigma_s = 1$, $\lambda_s = 0$, and
we return to the 2D solution given by Eq. (\ref{ER13}).

For a lattice of small absorbers of arbitrary shape on the reflecting
plane, an estimation of $\kappa$ can be deduced from the appeal to
electrostatic analogy (capacitance arguments), see Eq. (\ref{ER27}).
In fact, a given absorber can be replaced by an equivalent disk, which
has the same capacity as the absorber; after that, Eq. (\ref{ER28})
can be applied.  Unfortunately, this approach being physically
appealing has a limited range of validity, namely, a very diluted
limit (see \cite{Skvortsov_2023c} for details).  Another approach is
to use relation $K = 1/\mathcal{B}$, where $\mathcal{B}$ is the
blockage coefficient for a potential flow in a tube with a blocking
soft body of the shape of the protrusion.  For different shapes of
bodies, expressions for $\mathcal{B}$ are given in
\cite{Martin_2022}.

We also mention the perturbative computation of the trapping parameter
$\kappa$ for non-spherical axisymmetric boundaries with partial
reactivity $\kappa_0$ \cite{Piazza_2019}.  In sharp contrast to most
earlier studies dedicated to deformations of the plane, these
boundaries appear as small deformations of a sphere of radius $R$,
which can be written in spherical coordinates as $r(\theta) = R +
\delta f(\theta)$, where $\delta \ll R$, and $f(\theta)$ is a
dimensionless profile depending on the polar angle $\theta$.  If a
constant concentration $C_0$ is imposed at infinity, the diffusive
flux reads as
\begin{equation}  \label{eq:sphere_Japp}
J = \frac{4\pi D R C_0}{1 + D/(\kappa_0 R)} \biggl(1 + B_0 \frac{1 + \frac{2D}{\kappa_0 R}}{1 + \frac{D}{\kappa_0 R}} (\delta/R) 
+ O\bigl((\delta/R)^2\bigr)\biggr),
\end{equation}
where $B_0 = \tfrac12 \int\nolimits_0^\pi f(\theta) \sin(\theta)
d\theta$.  When $\delta = 0$, one retrieves the diffusive flux onto a
partially absorbing sphere found by Collins and Kimball
\cite{Collins_1949}, which is further reduced to the Smoluchowski flux
$J_0 = 4\pi D C_0 R$ for a perfectly absorbing sphere.  Comparison of
the diffusive fluxes for $\delta = 0$ and $\delta \ne 0$ allows one to
introduce the trapping parameter $\kappa$ of an effective spherical
boundary as:
\begin{equation}
\frac{1}{\kappa} = \frac{1}{\kappa_0}  - \biggl(\frac{1}{\kappa_0} + \frac{R}{D}\biggr) 
B_0 \frac{1 + \frac{2D}{\kappa_0 R}}{1 + \frac{D}{\kappa_0 R}} (\delta/R) + O\bigl((\delta/R)^2\bigr).
\end{equation}
This definition is meaningful only when the right-hand side is
positive.  One can also introduce the offset $\lambda$, even though
the original definition via Eq. (\ref{IE004}) has to be adapted to
this problem.

To conclude this section, we mention two examples of a coating made
from mixed absorbing/reflecting parts (reflecting protrusions on the
absorbing base).  For a hexagonal lattice of cylindrical pits of
height $h_0$ inside a reflecting matrix, one has
\cite{Dagdug_2003,Berezhkovskii_2011}
\begin{equation}
\label{ER41}
\frac{D}{\kappa} = \frac{h_0}{\sigma} + \frac{\sqrt{\mathcal{A}_0}}{F_1(\sigma)} ,
\end{equation}
where $\sigma$ is the absorbing fraction on the plane, $\mathcal{A}_0$
is the area of the periodic cell, and $F_1(\sigma)$ is given by
Eq. (\ref{ER30}).
In turn, for a hexagonal lattice of reflecting cylinders (pillars) on
the absorbing plane, the result reads
\begin{equation}
\label{ER42}
\frac{D}{\kappa}= \frac{h_0}{\sigma} +  \frac{\sqrt{\mathcal{A}_0}}{F_2(\sigma)} , 
\end{equation}
where $F_2(\sigma)$ is given by Eq. (\ref{ER33}).
The structure of Eqs. (\ref{ER41}), (\ref{ER42}) is very similar, with
two contributions.  The first term represents the diffusion resistance
of the pit, while the second term accounts for access resistance (a
delay due to the time needed for a particle to find an opening in the
interface).  The relative contributions of these terms depend on the
geometrical properties of the boundary profile, and the functional
form in the limits $h_0 \rightarrow 0$ and $\sigma
\rightarrow 0$ is very different.

\section{Constant-flux approximation}
\label{sec:constant}

In two previous sections, we mainly focused on absorbing spots or
stripes distributed periodically on the plane.  The same arguments and
often similar results can be derived for curved boundaries such as a
sphere (e.g., the Berg-Purcell's problem \cite{Berg_1977}) or a
cylinder (e.g., Stokes flow in a cylindrical pipe with stripes
\cite{Lauga_2003}), as discussed earlier.  Moreover, the periodicity
assumption can also be relaxed by considering random, uniformly
distributed arrangements of spots.  In fact, specific local variations
of the concentration near the absorbing spots are averaged out in the
far field, ensuring a similar form for the offset $\lambda$, up to
minor changes (in the same way as different lattices yielded different
constants in functions $F_1(\sigma)$ and $F_2(\sigma)$ in
Sec. \ref{sec:3D}).  At the same time, heterogeneous distribution of
absorbing spots or their polydispersity (i.e., distribution of their
sizes or shapes) can significantly alter the offset $\lambda$, as we
illustrated in Sec. \ref{sec:3D} by considering two complementary
configurations of absorbing patterns (Fig. \ref{fig:hexagon}a,b).  In
such situations, other approximations may be needed for estimating the
diffusive flux and the trapping parameter.

To give an example, we mention the so-called narrow escape problem
when a particle diffusing inside an Euclidean domain $\Omega \subset
\R^d$ with reflecting boundary $\pa$ searches for a small hole
$\Gamma$ located on that boundary \cite{Holcman_2013,Holcman_2014}.
Many efforts were dedicated to estimating the MFPT $T$ to this hole
\cite{Grigoriev_2002,Singer_2006a,Singer_2006b,Singer_2006c,Schuss_2007,Benichou_2008,Pillay_2010,Cheviakov_2010,Cheviakov_2012,Isaacson_2016,Dagdug_2016,Grebenkov_2016,Lindsay_2017,Bernoff_2018,Bernoff_2018b}.
To get accurate estimates in general domains, one needs to employ
elaborate mathematical tools such as conformal mapping (see
\cite{Holcman_2014,Grebenkov_2016} and references therein) or matched
asymptotic analysis (see \cite{Ward_1993} for details).  At the same
time, boundary homogenization can still be useful even in this setting
of a single small absorbing spot.  For instance, one of the first
results in this field is attributed to Lord Rayleigh who estimated $T
\approx |\Omega|/(4a D)$ for a small circular hole of radius $a$ on
the reflecting sphere of volume $|\Omega|$.  Curiously, this
leading-order result can be alternatively deduced by replacing the
small absorbing spot on the reflecting sphere by an effective
homogeneous partially absorbing sphere with the reactivity
$\kappa_{\rm BP}$ given by the Berg-Purcell's formula (\ref{ER35})
with $\mathcal{N} = 1$.  In this homogeneous setting, the MFPT can be
easily found as
\begin{equation}  \label{eq:Tsphere}
\overline{T} = \frac{R^2}{15D} + \frac{R}{3\kappa_{\rm BP}}
\end{equation}
after averaging over the starting point.  Substituting $\kappa_{\rm
BP}$ from Eq. (\ref{ER35}), one notices that the second term in
Eq. (\ref{eq:Tsphere}) is equal to $|\Omega|/(4a D)$ and provides the
dominant contribution if $a \ll R$.  In other words, despite highly
non-uniform distribution of absorbing spots, boundary homogenization
still yields the correct result in the leading order.  In turn, more
advanced techniques are needed for getting next-order corrections or
analyzing the dependence on the starting point.

In this section, we briefly describe another approximation that allows
to deal with non-uniform distributions of absorbing spots.  The
so-called constant-flux approximation (also known as self-consistent
approximation) was proposed by Shoup {\it et al.} to calculate the
Smoluchowski-type constant for reactions with a small center situated
on the otherwise impenetrable surface of a spherical domain
\cite{Shoup_1981} (see also earlier works by Keller and Stein
\cite{Keller_1967}).  Let us formulate it in a more general setting of
diffusion in an unbounded domain $\Omega$, whose reflecting boundary
$\pa$ contains partially absorbing spots $\Gamma$ characterized by
reactivity $\kappa_0$.  As previously, the steady-state concentration
$C$ obeys the Laplace equation, $\Delta C = 0$, with a constant
concentration $C_0$ at infinity, and mixed boundary conditions:
\begin{subequations}  \label{eq:constant_mixed}
\begin{align}  \label{eq:constant_Robin}
- D\partial_n C & = \kappa_0 C  \quad \textrm{on}~\Gamma,  \\ 
- D\partial_n C & = 0 \quad \textrm{on}~ \pa\backslash \Gamma.
\end{align}
\end{subequations}
The reactivity $\kappa_0$ can either represent finite reaction
probability on the spots, or result from the previous step of boundary
homogenization on $\Gamma$.  Note that the limit $\kappa_0 = \infty$
corresponds to the absorbing spots discussed in Sec. \ref{sec:3D}.  An
exact solution of this boundary value problem is difficult even for
simple domains such as a disk or a sphere.  In particular, even for
planar domains, conformal mapping is not much useful for finite
$\kappa_0$ because such a transformation will make the reactivity
dependent on the boundary point.  The basic idea of the constant-flux
approximation consists in replacing the right-hand side of Robin
boundary condition (\ref{eq:constant_Robin}) by a constant flux
density $j_0$.  In other words, one aims at approximating the solution
$C$ of the original problem by the solution $\hat{C}$ of the modified
problem, $\Delta
\hat{C} = 0$, with inhomogeneous Neumann boundary condition:
\begin{equation}
-D\partial_n \hat{C} = j_0 \, \mathbb{I}_\Gamma(\x)  \quad \textrm{on}~\pa,
\end{equation}
where $\mathbb{I}_\Gamma(\x)$ is the indicator function of the spots:
$\mathbb{I}_\Gamma(\x) = 1$ for $\x\in\Gamma$, and $0$ otherwise.  The
unknown parameter $j_0$ can be fixed by imposing the self-consistent
condition that the Robin boundary condition (\ref{eq:constant_Robin})
is satisfied {\it on average} on $\Gamma$:
\begin{equation}
\kappa_0 \int\limits_{\Gamma} \hat{C}(\x) d\x = \int\limits_{\Gamma} \bigl(- D\partial_n \hat{C}\bigr) d\x = j_0 |\Gamma| ,
\end{equation}  
where $|\Gamma|$ is the surface area of the spots $\Gamma$.  Since
$\hat{C}(\x)$ implicitly depends on $j_0$, this equation can be used
to determine $j_0$ (in the limit $\kappa_0 = \infty$, this equation is
reduced to $\int\nolimits_{\Gamma} \hat{C}(\x) d\x = 0$, which still
determines $j_0$).  Even though finding the solution $\hat{C}$ of the
modified problem is in general not elementary, it is much simpler than
solving the original problem with mixed boundary conditions
(\ref{eq:constant_mixed}).

Since the diffusive flux density of the exact solution, $-D\partial_n
C$, is not constant, this approximation fails in a vicinity of the
absorbing spot, in the same way as the far-field approximation
(\ref{ER2}) failed near the boundary.  However, the approximate
solution $\hat{C}$ turns out to be accurate far from the absorbing
spots.  In particular, the approximation $J = j_0 |\Gamma|$ for the
diffusive flux is very accurate in many settings.  In this light, the
constant-flux approximation can be seen as an alternative way of
boundary homogenization discussed in sections \ref{sec:2D} and
\ref{sec:3D}.  The major advantage of the constant-flux approximation
is that the diffusive flux density is not uniformly distributed over
the boundary $\pa$, as it was for homogenized boundaries.  The
accuracy of two methods was compared for absorbing stripes on a
cylinder \cite{Grebenkov_2019a}.

The original constant-flux approximation by Shoup {\it et al.} was
later extended and applied to other problems such chemical kinetics of
active colloidal particles \cite{Oshanin_2017}, MFPT in cylindrical
and spherical domains \cite{Grebenkov_2017a,Grebenkov_2017b}, and the
whole distribution of first-passage times
\cite{Grebenkov_2018,Grebenkov_2019,Grebenkov_2021}.  For instance,
for diffusion between two capped coaxial cylinders of height $H$ and
radii $\rho$ and $R$ (Fig. \ref{fig:cylinders}e), the MFPT to an
absorbing stripe located on the inner cylinder was found in
\cite{Grebenkov_2017b}.  When the starting point is uniformly
distributed, the constant-flux approximation yields 
\begin{equation}  \label{eq:T_SCA}
T = \frac{H(R^2 - \rho^2)}{\pi D \rho} f,  \qquad 
f = \sum\limits_{n=1}^\infty \frac{G_n}{n} \biggl(\frac{\sin(\pi \sigma n)}{\pi \sigma n}\biggr)^2,
\end{equation}
with 
\begin{equation}
G_n = -\frac{I_1(\pi n R/H) K_0(\pi n \rho/H) + K_1(\pi n R/H) I_0(\pi n \rho/H) } 
{I_1(\pi n R/H) K_1(\pi n \rho/H) - K_1(\pi n R/H) I_1(\pi n \rho/H) } ,
\end{equation}
where $I_\nu(z)$ and $K_\nu(z)$ are the modified Bessel functions of
the first and second kind.  In turn, using Eq. (\ref{ER8}) for the
trapping parameter $\kappa$ (with the correction factor $1/2$ due the
perpendicular orientation of stripes, see \cite{Lauga_2003}), one gets
a much simpler approximation $f \approx -\ln(\sin(\pi \sigma/2))$.
While the simplicity of this expression is advantageous for getting
rapid estimates, the constant-flux approximation in
Eq. (\ref{eq:T_SCA}) is more accurate \cite{Grebenkov_2019a}.

Another example of the fruitful use of the constant-flux approximation
is the computation of the steady-state reaction rate of
diffusion-controlled reactions in sheets \cite{Grebenkov_2018b}.  In
some applications, the absorbing spot is located on one side of a flat
layer (a sheet) between two parallel reflecting boundaries and can
accessed from the lateral sides (Fig. \ref{fig:cylinders}d).  For
instance, one can think of a capped cylinder $\{ (x,y,z)\in\R^3~:~ x^2
+ y^2 < R^2,~ 0 < z < H\}$ of height $H$ and radius $R$, with a
constant concentration imposed on the cylinder wall and an absorbing
disk of radius $\rho$ lying on the bottom.  A semi-analytical solution
of the Laplace equation, its constant-flux approximation and
asymptotic behavior were derived in \cite{Grebenkov_2018b}.  In
particular, the constant-flux approximation yielded the following
asymptotic behavior
\begin{equation}
\frac{J}{J_0} \approx \frac{\pi}{2} \biggl(\frac{16}{3\pi} + \frac{\rho}{H} \ln \frac{R}{2H}\biggr)^{-1}  \qquad (\rho \ll H \ll R),
\end{equation}
where $J_0 = 4\rho D C_0$ is the diffusive flux on the half-disk in
the upper half-space (with $4\rho$ being the half of the capacity of
the disk).  This ratio determines the offset $\lambda$ and the
trapping parameter $\kappa$.  Other asymptotic limits were discussed
in \cite{Grebenkov_2018b}.

\section{Diffusion inside channels}
\label{sec:channel}

\begin{figure}[t!]
\centering
\includegraphics[width=0.29\textwidth]{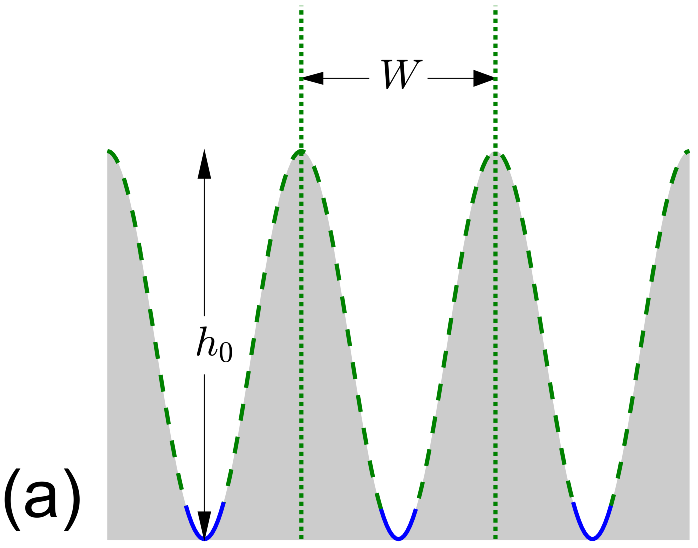}  \hskip 3mm
\includegraphics[width=0.36\textwidth]{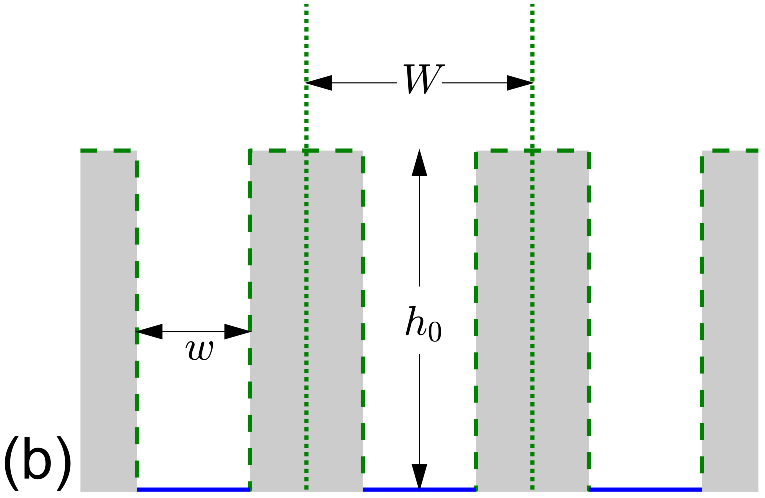}  \hskip 3mm
\includegraphics[width=0.28\textwidth]{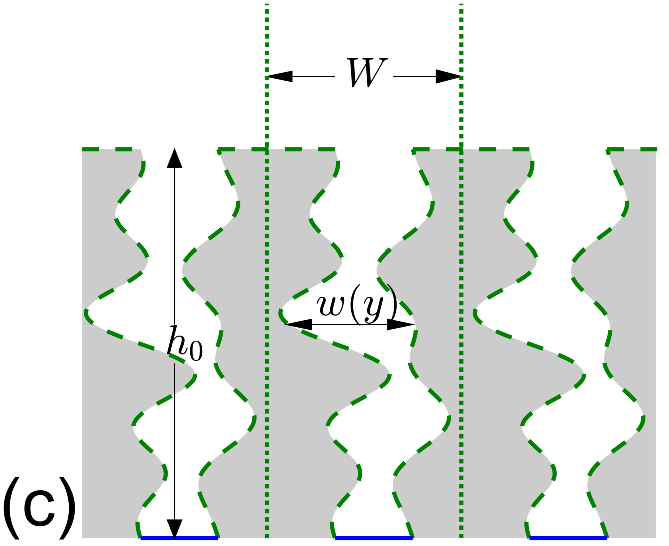}
\caption{
Arrays of long pits with buried binding sites at the bottom (in blue).
{\bf (a)} Sinusoidal profile.  {\bf (b)} Rectangular profile of width
$w$ and height $h_0$, on the period $W$, with a flat absorbing site at
the bottom.  {\bf (c)} Smoothly varying profile $w(y)$ of a pit of
height $h_0$, with a flat absorbing site at the bottom.  Light gray
indicates inaccessible solid matrix.  The particles arrive from a flat
source at $y = H$ located far above the entrances of the pits.
Vertical dotted lines delimit one periodic cell.}
\label{fig:pits}
\end{figure}

Finding buried binding sites by diffusing particles can be seen as
another variant of target search problems
\cite{Samson_1978,Dagdug_2003,Berezhkovskii_2011}.  In this problem,
the binding chemical sites (usually modeled as a finite domain with
absorbing boundary) are at the bottom of the deep pits made inside a
reflecting porous matrix or at the bottom of a deep reflecting canopy
(Fig. \ref{fig:pits}a).  The wall of the pit creates the geometrical
constraint (an entropy barrier) preventing particles to access the
absorbing bottom.  This setting is an extension of the example of a
semi-circular canopy shown in Fig. \ref{fig:profiles}c.  In this
section, we briefly discuss a powerful approximation via the
Fick-Jacobs equation that allows one to reduce the original 2D or 3D
problems to an effective one-dimensional problem that can be solved
explicitly
\cite{Kalinay_2005,Kalinay_2006,Kalinay_2008,Mangeat_2017,Mangeat_2018,Berezhkovskii_2007,Berezhkovskii_2015}.
We start with the planar setting and then briefly mention the 3D case,
which is very similar.

\subsection*{Offset and trapping parameter of a pit} 

Let us consider a periodic array of 2D deep pits with reflecting walls
and the absorbing bottom (Fig. \ref{fig:pits}b,c).  As previously, one
can focus on a single periodic cell of width $W$.  The Fick-Jacobs
equation inside the pit has the form
\cite{Kalinay_2005,Kalinay_2006,Kalinay_2008,Mangeat_2017,Mangeat_2018}
\begin{equation}
\label{ER23}
 \frac{d}{dy} \left[ D(y) w(y) \frac{d}{dy} \frac{c(y)}{w(y)}  \right]=0,  ~~~ 0 < y < h_0, 
\end{equation} 
where $w(y)$ is the width of the pit at height $y$, $c(y)$ is the
effective one-dimensional concentration, and $D(y)$ is the effective
diffusivity determined by $w(y)$.  Among numerous approximations for
$D(y)$, the model of Kalinay and Percus,
\begin{equation}
\label{ER24}
D(y) =D \frac{\atan(\frac{1}{2} w'(y))}{\frac{1}{2} w'(y)},
\end{equation}
is often employed \cite{Kalinay_2005,Kalinay_2006,Kalinay_2008}.  For
a smooth profile ($w'(y) \ll 1 $), we assume $D(y) = D=const$ in a
first approximation \cite{Berezhkovskii_2015}.  The absorbing bottom
of the pit implies the boundary condition $c=0$, whereas we set $c=c_0
= const$ at some $y = H \gg W$ at the far field.
The offset $\lambda$, which is determined by the far-field behavior
(\ref{ER2}) at large $y$, is then evaluated from the explicit solution
of Eq. (\ref{ER23}) as
\begin{equation}
\label{ER25}
\lambda = h_0 - W \int^{h_0}_0\frac{dy}{w(y)} .
\end{equation}
Setting an effective radiative boundary at $y_0 = h_0$, one gets the
trapping parameter
\begin{equation}
\label{ER25_bis}
\kappa = \frac{D}{W} \left(\int^{h_0}_0\frac{dy}{w(y)} \right)^{-1} .
\end{equation}

If there are discontinuities in the profile $w(y)$ (e.g., near the pit
entrance, $y = h_0$, as illustrated in Fig. \ref{fig:pits}b,c), then
the Fick-Jacobs equation is not applicable near these points.  In this
case, one first employs boundary homogenization to estimate the
trapping parameter $\kappa$ in the 2D periodic cell, and then imposes
a transmission boundary condition for the one-dimensional
concentration $c(y)$ (see
\cite{Kalinay_2005,Kalinay_2006,Kalinay_2008,Mangeat_2017,Mangeat_2018}
and references therein).  This boundary condition results in the
additional term in the formula (\ref{ER25_bis}) for $\kappa$, see
\cite{Dagdug_2021}.  The physical nature of this term (the so-called
access resistance) is a delay in particle diffusion due to the time
taken by the particle to find the pit entrance.

The above analysis can be carried on in three dimensions for a lattice
of long pits by replacing the pit width $w(y)$ by its cross-sectional
area $\mathcal{A}(y)$ that leads to a general formula
\begin{equation}
\label{ER43}
\lambda = h_0 - \mathcal{A}_0 \int^{h_0}_0\frac{dy}{\mathcal{A}(y)},
\end{equation} 
where $\mathcal{A}_0 $ is the area of the periodic cell.  In this way,
the trapping parameter $\kappa$ can be analytically derived for a
variety of pit profiles (cone, spheroid, spindle, see
Refs. \cite{Skvortsov_2023d,Skvortsov_2023b,Dagdug_2022a} for
details).  We note that the change of the pit profile can modify the
functional form of $\kappa(\sigma)$ \cite{Martin_2022}.
Similar to Eq. (\ref{ER41}), the discontinuity in the pit profile near
the opening leads to an additional contribution to $\kappa$
\cite{Skvortsov_2023d,Skvortsov_2023b,Dagdug_2022a}.

\subsection*{Mean first-passage time}

\begin{figure}[t!]
\centering
\includegraphics[width=0.5\textwidth]{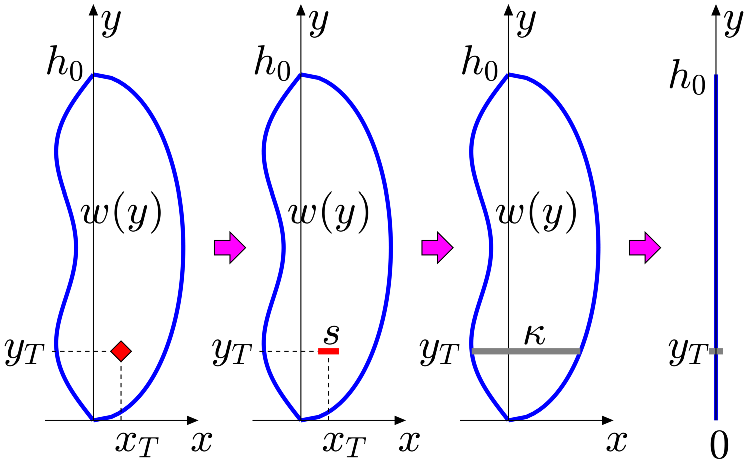} 
\caption{
Three-step transformation of the original MFPT problem inside an
elongated planar domain (left) into an effectively one-dimensional
problem with semi-permeable semi-absorbing boundary condition (right).
A small target of arbitrary shape (here, a red rhombus) is replaced by
an equivalent interval (in red) of length $s$ of the same
log-capacity, which is then replaced by a semi-permeable
semi-absorbing boundary (in gray) with reactivity $\kappa$.}
\label{fig:profiles2D}
\end{figure}

In a similar way, the Fick-Jacobs equation can be applied to estimate
the MFPT to a small target inside an 2D elongated domain with
reflecting boundaries \cite{Grebenkov_2020}.  The profile $w(y)$ of
the domain is assumed to be smooth, slowly changing, but otherwise
general, whereas the target is small but of an arbitrary shape.
To treat the original 2D problem as one-dimensional
(Fig. \ref{fig:profiles2D}), we first replace the absorbing target by
an equivalent absorbing interval (i.e., the interval of the same
log-capacity).  In turn, this absorbing interval is then replaced by a
partially absorbing horizontal segment of length $w(y_T)$ and trapping
parameter $\kappa$.  The latter can be estimated from Eq. (\ref{ER9})
by thinking of the elongated domain as a stripe of width $w(y_T)$ with
reflecting boundaries.  We stress that the value of the trapping
parameter $\kappa$ depends on the position $y_T$ of the target that
enables formulating the MFPT within the effective medium
approximation.  After that, the Poisson equation for MFPT in the
elongated domain reduces to 1D equation that resembles the Fick-Jacobs
equation:
\begin{equation}
\label{ER44}
 \frac{D}{w(y)} \frac{d}{dy} \left[ w(y)\frac{d \mathcal{T}}{dy} \right] = -1,
\end{equation}
subject to the effective semi-permeable boundary condition at the
target location $y_T$: 
\begin{align}
\label{ER45}
 \mathcal{T} (y_{T} - 0) & = \mathcal{T} (y_{T} + 0),  \\
\label{ER46}
 D \left[\frac{d \mathcal{T}}{dy}  (y_{T} + 0) -  \frac{d \mathcal{T}}{dy}  (y_{T} - 0) \right] & = \kappa\, \mathcal{T} (y_{T}) .
\end{align}
This problem was then solved analytically for a variety of profiles
$w(y)$ and favorably validated by numerical simulations
\cite{Grebenkov_2020}.

This framework has been extended to find the MFPT in 3D elongated
domains with axial symmetry \cite{Grebenkov_2022}.  Here,
Eq. (\ref{ER44}) remains the same, with $w(y)$ being replaced by a
cross-sectional area $\mathcal{A}(y)$ of the domain.  The target is
modeled by an equivalent disk and the trapping parameter $\kappa$ is
given by Eq. (\ref{ER28}) via boundary homogenization.  We stress that
the ideas of boundary homogenization can provide some analytical tools
not only for calculating the MFPT, but also for evaluating the FPT
statistics, in particular, in the long-time limt
\cite{Grebenkov_2023}.

\section{Conclusion}
\label{sec:conclusion}

Laplacian transport near complex boundaries is relevant in many
disciplines, including chemical physics, electrostatics,
hydrodynamics, and heat transfer.  Despite the long history of
intensive research, this field remains very active, with numerous
recent advances and applications.  In this review, we overiewed
several approximations that yielded a simplified yet accurate
description of this phenomenon.  In particular, we discussed the
effective medium theory, boundary homogenization, constant-flux
approximation and perturbative analysis that allow one to express the
diffusive flux and other integral properties of the system in terms of
its geometric and absorption parameters.  For instance, we described
the procedure of boundary homogenization when the complex boundary is
replaced by an equivalent flat boundary with Robin boundary condition.
The fraction of absorbing parts and the amplitude of boundary profile
variations do not need to be small, while the boundary can be rough
and spiky.  The proposed framework may be useful for a rapid modeling
and engineering prototyping before proceeding with extensive computer
simulations.

At the same time, many other aspects of Laplacian transport remain
unexplored in this review: (i) the narrow escape problem and the
related matched asymptotic analysis; (ii) diffusive search in porous
media formed by absorbing/reflecting spherical beads and efficient
semi-analytical methods
\cite{Galanti_2016,Grebenkov_2019b,Grebenkov_2020b}; (iii) Laplacian
transport towards fractal boundaries
\cite{Witten_1981,Mandelbrot_1990,Filoche_2000,Grebenkov_2005b,Levitz_2006,Andrade_2007,Filoche_2008,Rozanova_2012}
and its electrochemical measurements
\cite{deLevie_1965,Liu_1985,Halsey_1987,deLevie_1990,Halsey_1992,Pajkossy_1994,Chassaing_1994};
(iv) the impact of partial reactivity of complex boundaries, its
approximate treatments (including land surveyor approximation
\cite{Sapoval_1994,Sapoval_1996,Sapoval_2001}), and the spectral theory based on the
Dirichlet-to-Neumann operator
\cite{Grebenkov_2006,Grebenkov_2006b,Grebenkov_2015,Grebenkov_2019d,Grebenkov_2020c};
(v) time-dependent (non-stationary) diffusion in complex media, the
distribution of first-passage times, and efficient search strategies
\cite{Benichou_2011,Benichou_2014,Metzler_2014,Grebenkov_2018e,Lindenberg_2019}.
We believe that further applications of boundary homogenization and
other approximate tools can help uncovering the crucial role of
complex geometry and heterogeneity in Laplacian transport and target
search problems.

\section*{Acknowledgments}

A.T.S thanks Alexander M. Berezhkovskii and Paul A. Martin for useful
discussions.

\end{document}